\documentclass[lettersize,journal]{IEEEtran}
\usepackage{amsmath,amsfonts}
\usepackage{booktabs}
\usepackage{algorithmic}
\usepackage{algorithm}
\usepackage{array}
\usepackage[caption=false,font=normalsize,textfont=sf]{subfig}
\usepackage{textcomp}
\usepackage{stfloats}
\usepackage{url}
\usepackage{verbatim}
\usepackage{graphicx}
\usepackage{cite}
\usepackage{multicol}
\usepackage{amssymb}
\usepackage{lineno}
\usepackage{makecell}
\usepackage{indentfirst}
\usepackage{newtxmath}
\usepackage{color}
\usepackage[colorlinks,
linkcolor=blue,
anchorcolor=blue,
citecolor=blue]{hyperref}

\usepackage{cite}
\usepackage{epstopdf}
\usepackage{xcolor}
\usepackage{tikz}
\usepackage{mathtools}
\usepackage{pifont} 
\usepackage{tikz}
\usetikzlibrary{positioning}
\newcommand{\diag}{\mathrm{diag}}

\newcommand{\tr}{\mathrm{tr}}
\newcommand{\HH}{\mathrm{H}}	
\newcommand{\TT}{\mathrm{T}}

\newtheorem{theorem}{\textbf{Theorem}}

\newtheorem{remark}{\textbf{Remark}}
\newtheorem{corollary}{\textbf{Corollary}}
\newtheorem{proposition}{\textbf{Proposition}}
\newtheorem{assumption}{\textbf{Assumption}}
\newtheorem{lemma}{\textbf{Lemma}}

\begin{document}
	
	\title{Adaptive Matched Filtering for Sensing With Communication Signals in Cluttered Environments}

	\author{Lei Xie,~\IEEEmembership{Member,~IEEE}, Hengtao He,~\IEEEmembership{Member,~IEEE}, Yifeng Xiong,~\IEEEmembership{Member,~IEEE},\\ Fan Liu,~\IEEEmembership{Senior Member,~IEEE},
		and Shi Jin,~\IEEEmembership{Fellow,~IEEE}
		\thanks{L. Xie is with School of Cyber Science and Engineering, Southeast University, Nanjing 210096, China.} 
		\thanks{H. He, F. Liu, and S. Jin are with School of Information Science and Engineering, Southeast University, Nanjing 210096, China. \emph{(Corresponding author: Fan Liu.)}} 
		\thanks{Y. Xiong is with School of Information and Electronic Engineering, Beijing University of Posts and Telecommunications, Beijing 100876, China.}
	}

	\maketitle
	
	\begin{abstract}
		This paper investigates the performance of the adaptive matched filtering (AMF) in cluttered environments, particularly when operating with superimposed signals. Since the instantaneous signal-to-clutter-plus-noise ratio (SCNR) is a random variable dependent on the data payload, using it directly as a design objective poses severe practical challenges, such as prohibitive computational burdens and signaling overhead. To address this, we propose shifting the optimization objective from an instantaneous to a statistical metric, which focuses on maximizing the average SCNR over all possible payloads. Due to its analytical intractability, we leverage tools from random matrix theory (RMT) to derive an asymptotic approximation for the average SCNR, which remains accurate even in moderate-dimensional regimes.
		A key finding from our theoretical analysis is that, for a fixed modulation basis, the phase-shift keying (PSK) achieves a superior average SCNR compared to quadrature amplitude modulation (QAM) and the pure Gaussian constellation. Furthermore, for any given constellation, the orthogonal frequency division multiplexing (OFDM) achieves a higher average SCNR than single-carrier (SC) and affine orthogonal frequency-division multiplexing (AFDM).
		Then, we propose two pilot design schemes to enhance system performance: a Data-Payload-Dependent (DPD) scheme and a Data-Payload-Independent (DPI) scheme. The DPD approach maximizes the instantaneous SCNR for each transmission. Conversely, the DPI scheme optimizes the average SCNR, offering a flexible trade-off between sensing performance and implementation complexity. Then, we develop two dedicated optimization algorithms for DPD and DPI schemes. In particular, for the DPD problem, we employ fractional optimization and the Karush–Kuhn–Tucker (KKT) conditions to derive a closed-form solution. For the DPI problem, we adopt a manifold optimization approach to handle the inherent rank-one constraint efficiently. Simulation results validate the accuracy of our theoretical analysis and demonstrate the effectiveness of the proposed methods.
	\end{abstract}
	
	
	\section{Introduction}
	The six-generation (6G) communication system has emerged as a transformative platform that will support advanced applications such as autonomous vehicles, smart factories, digital twins, and the low‑altitude economy \cite{liu2022integrated,yuan2025ground,wu2025low}. 
	Among the enabling technologies for 6G, integrated sensing and communication (ISAC) has been recognized as a particularly promising solution \cite{jiang2025integrated}. Within the ISAC framework, the same hardware and wireless resources are exploited to achieve both high-rate data transmission and high-precision environmental sensing, supporting tasks such as target detection, tracking, and imaging \cite{liu2018toward,liu2020joint,xie2023collaborative}. Existing studies on ISAC system design have mainly focused on analyzing the effect of random communication signals on sensing performance \cite{xiong2023fundamental,lu2023random,xie2023sensing}. While these works provide valuable insights into average estimation performance, they are generally unable to characterize the fundamental performance of target detection.
	
	A key challenge for target detection is that targets typically exhibit “low, slow, and small” characteristics, i.e., low altitude, slow speed, and a small radar cross-section \cite{9362227}. These attributes pose significant challenges for reliable detection \cite{wang2024low}. To extract target information from weak echo signals, matched filtering (MF) is commonly employed to maximize the output signal-to-noise ratio (SNR) and thereby enhance detection performance \cite{1057571,leuschen2002matched}. 
	To characterize the performance of MF with random ISAC signals, the radar ambiguity function of ISAC waveforms has been analyzed in recent studies \cite{liu2025cp,liu2025uncovering}.
	In particular, the authors of \cite{liu2025cp}  provided a powerful theoretical foundation: when designing communication-centric ISAC systems using standard quadrature amplitude modulation (QAM) or phase-shift keying (PSK) constellations, the orthogonal frequency division multiplexing (OFDM) is the optimal choice not only for its excellent performance in anti-multipath communications but also for its superior ranging performance.
	In \cite{liu2025uncovering}, a fundamental analysis of the sensing performance of communication-centric ISAC signals was provided, which extends previous work by incorporating the crucial effects of pulse shaping. It confirms the optimality of OFDM for standard communication constellations and introduces a novel “iceberg shaping” design approach that allows for designing ISAC signals with superior sidelobe suppression. While existing studies have established an important foundation for understanding how random communication signals affect sensing performance, several key challenges must be overcome to fully unleash the potential of ISAC. In particular, two key challenges emerge: the sub-optimality of conventional detectors in cluttered environments and the hybrid nature of realistic ISAC signals.

	First, although MF is considered the optimal detector in an ideal additive white Gaussian noise environment, it may not fit in the cluttered scenarios. 
	In particular, the environment is often saturated with strong clutter echoes, which are typically highly correlated in the spatial and/or temporal domains with the target. When such correlated clutter is processed by a conventional matched filter, effective suppression can not be achieved. This results in a significant residual clutter component at the output of MF in the test cell. In high-clutter regions, the residual clutter component can overwhelm true targets, which may result in a high increase in the false alarm; while in low-clutter regions, targets may be missed due to an excessively high threshold. To address this limitation, the adaptive matched filtering (AMF) was proposed \cite{fuhrmann1992cfar,roman2000parametric,qureshi2018parametric}. Unlike the MF, the AMF can dynamically adjust its filter weight to adapt to the real-time cluttered environment \cite{6166358,6853408,xie2020recursive}. Specifically, the received data is “whitened” to decorrelate the clutter and transform it into a white-noise-like process, thereby achieving effective clutter suppression. Then, a matched filter is applied to this whitened data to maximize the output signal-to-clutter-plus-noise ratio (SCNR). This adaptive approach significantly enhances the ability to detect weak targets in strong clutter environments \cite{1337465,conte2002adaptive,blum2002analysis}.
	
	Second, most existing studies focus on the assumption of purely random signals \cite{xiong2023fundamental,lu2023random,xie2023sensing}. However, in 5G New Radio (NR) systems, the transmitted signals are neither entirely deterministic nor purely random. In a typical NR frame structure, pilots occupy about 10$\%$ of the time/frequency resources, while the remaining majority, which comprises random data payloads, can be exploited for sensing purposes \cite{lin20215g}. Consequently, recent works have investigated hybrid ISAC waveforms that combine deterministic pilot symbols with random data payloads. 
	In \cite{xie2025bistatic}, the paper validates the effectiveness of the proposed generalized likelihood ratio test (GLRT) detector and demonstrates its superior performance over conventional methods that use only pilots for detection, especially when the number of samples is limited. This underscores the necessity of designing dedicated detectors for hybrid ISAC signals. However, this study primarily focuses on the conventional embedded pilot scheme, in which the pilot and data payload are orthogonal to each other in either the time or frequency domain \cite{4014391}. While such orthogonality simplifies the processing, it inevitably incurs a loss of spectral efficiency. To overcome this limitation, the concept of the superimposed pilot has been proposed \cite{upadhya2017superimposed,zhang2025afdm}, where the pilot signal is overlaid onto the data payload to enable simultaneous transmission. Nevertheless, the sensing performance associated with superimposed pilots and data payloads remains largely unexplored. 
	
	This paper investigates the AMF in cluttered environments with communication signals. Unlike previous studies that consider pilot and data payload separately, we jointly exploit both for sensing. In particular, we focus on two critical and open research questions:
	1) How to evaluate the performance of AMF with communication signals; and 2) How to optimize the performance through pilot design. To achieve the first objective, we first derive the instantaneous SCNR. Since it depends on the random data payload, the instantaneous SCNR itself becomes a random variable, rendering per-slot optimization impractical due to its excessive computational and signaling overhead. To overcome this limitation, we reformulate the design objective by focusing on the average SCNR. Due to its analytical intractability, we derive an asymptotic approximation of the average SCNR based on tools from random matrix theory. Then, we propose two pilot design schemes, i.e., data-payload-dependent (DPD) and data-payload-independent (DPI), which maximize the instantaneous and average SCNR, respectively. Simulation results validate the accuracy of the theoretical analysis results and the effectiveness of the proposed DPD and DPI methods.
	
	\begin{figure}[!t]
		\centering
		\includegraphics[width=3.5in]{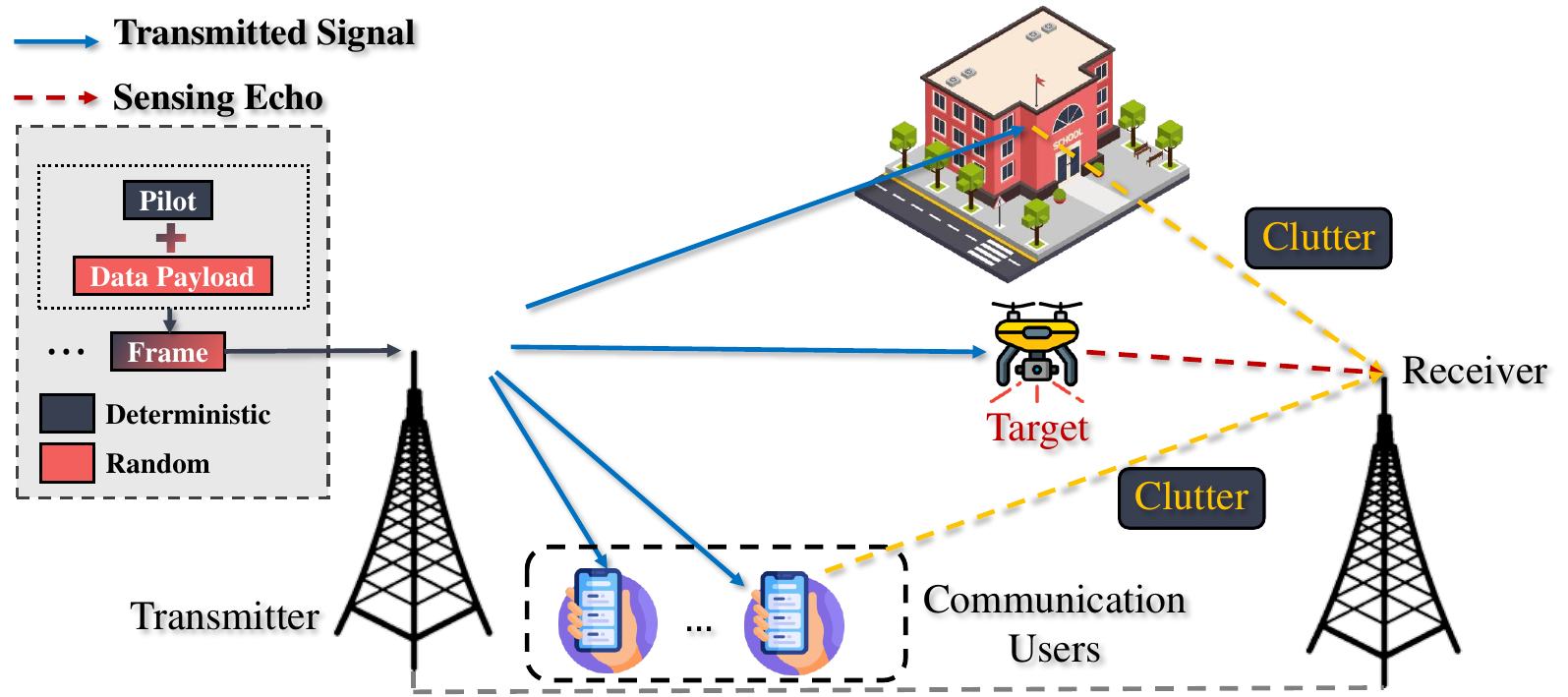}
		\caption{Illustration of the considered ISAC system}
		\label{Ill_sysyem}
	\end{figure}
	
	The main contributions of this paper are summarized as:

	\begin{enumerate}
		\item We design the AMF for communication signals. By employing tools from random matrix theory, we derive explicit expressions for the average SCNR. Though the expression is derived in the infinite‑dimensional regime, it is still accurate in the moderate condition. Our results show, in terms of the average SCNR, that PSK is better than both QAM and the pure Gaussian constellation, whereas OFDM is better than single‑carrier (SC) transmission and affine orthogonal frequency‑division multiplexing (AFDM).
		\item Building upon the theoretical results, we propose two pilot design methods, which are referred to as DPD and DPI, to maximize the instantaneous and average SCNR, respectively. In particular, DPD designs the pilot based on the transmitted data payload to maximize the instantaneous SCNR for each transmission. DPI focuses on maximizing the average SCNR over all possible payloads. The two methods enable a trade-off between sensing performance and implementation complexity. 
		\item To solve the DPD and DPI design problems, we develop two dedicated algorithms. Specifically, for the DPD problem, we first employ fractional optimization to reformulate the original non-convex problem into an equivalent convex form that can be efficiently solved. Then, we apply the Karush–Kuhn–Tucker (KKT) conditions to derive a closed-form update for the optimization variable. For the DPI problem, we adopt a manifold optimization approach to effectively handle the inherent rank-one constraint, which enables efficient convergence to a high-quality solution.
	\end{enumerate}

	The remainder of this paper is organized as follows. Sec. II presents the system model. Sec. III derives an explicit expression for the average SCNR and provides corresponding physical insights. Sec. IV introduces two pilot optimization approaches aimed at maximizing the instantaneous and average SCNR, respectively. Sec. V presents simulation results that validate the theoretical analysis and demonstrate the effectiveness of the proposed methods. Finally, Sec. VI concludes the paper and summarizes the key findings and contributions.

	\section{System Model}

	Consider an ISAC system composed of a single-antenna transmitter and receiver, as shown in Fig. \ref{Ill_sysyem}. In this paper, we consider the simultaneous downlink transmission and target sensing with the bi-static sensing scheme. In the following, we will present the transmitted and received models, respectively.

	\subsection{Transmitted Signal}

	\begin{figure}[!t]
		\centering
		\includegraphics[width=3.5in]{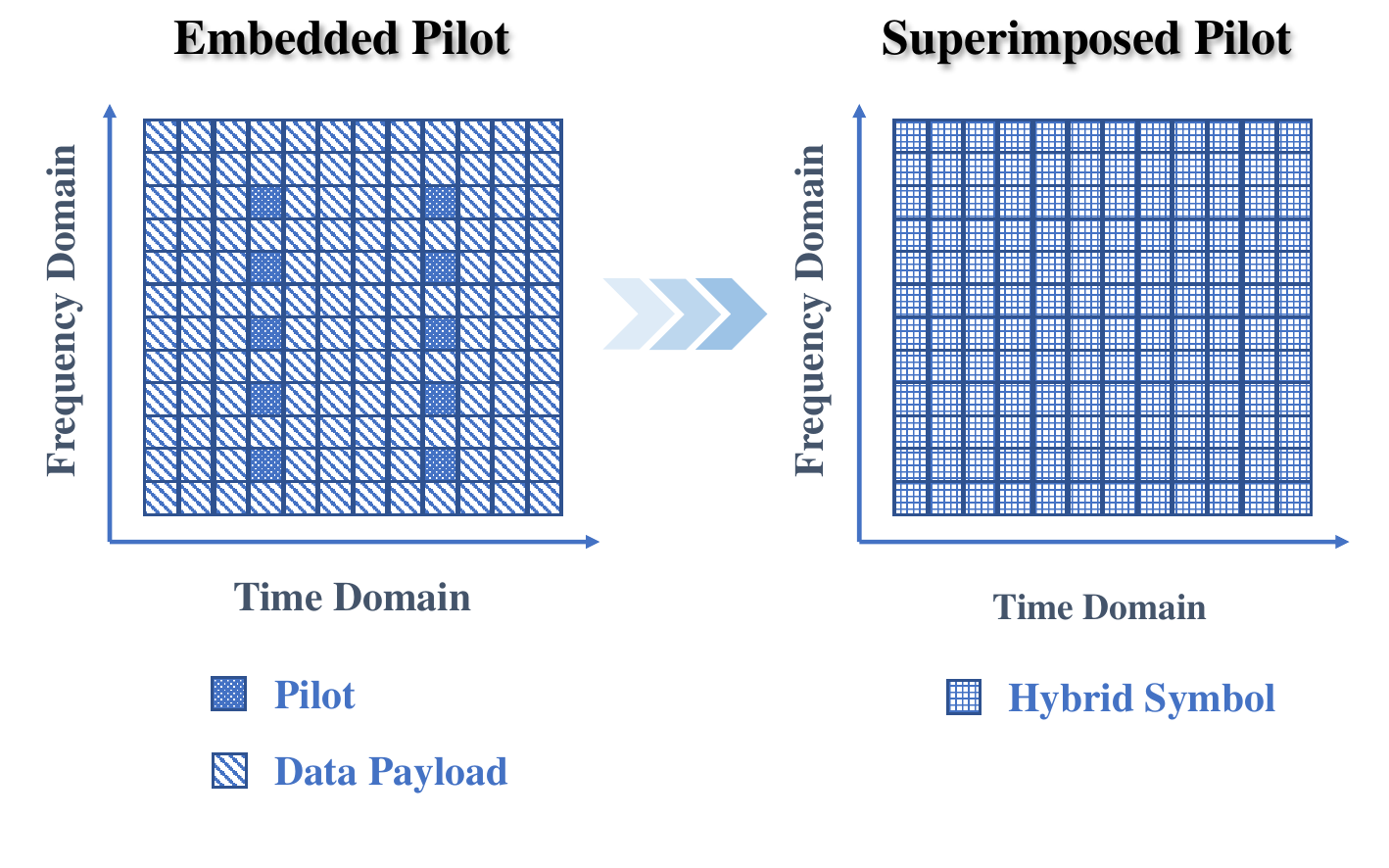}
		\caption{Illustration of the superimposed pilot scheme}
		\label{Ill_Superimposed_Pilot}
	\end{figure}

	In this paper, we consider a superimposed pilot scheme, as illustrated in Fig. \ref{Ill_Superimposed_Pilot}. Unlike the conventional embedded pilot approach, where dedicated time or frequency resources are reserved exclusively for pilot transmission, the superimposed pilot embeds the pilot signal directly onto the data symbols. This design improves spectral efficiency (SE), as both pilot and data share the same transmission resources. 
	In particular, the transmitted signal is modeled by
	\begin{equation}
		\mathbf{x} = \mathbf{x}_p + \mathbf{x}_d \in \mathcal{C}^{N\times 1},
	\end{equation}
	where $\mathbf{x}_p$ denotes the deterministic pilot and $\mathbf{x}_d$ denotes the random data payload. 
	
	In typical communication systems, a block of $N$ symbols is modulated over an orthonormal modulation basis defined in the time domain. Such a basis can be compactly represented by a unitary matrix $\mathbf{U} \in \mathbb{U}(N)$, where $\mathbb{U}(N)$ denotes the unitary group of degree $N$, i.e., the set of all $N \times N$ unitary matrices satisfying $\mathbf{U}^\mathsf{H}\mathbf{U} = \mathbf{I}_{N}$. The adoption of a unitary basis guarantees that the modulation process is energy-preserving, while also ensuring that the transmitted symbols remain mutually orthogonal. Consequently, letting $\mathbf{s}_d \in \mathbb{C}^{N\times 1}$ denote the vector of data symbols, the corresponding discrete time-domain transmit signal is given by \cite{liu2025cp}
	\begin{equation}
		\mathbf{x}_d = \mathbf{U}\mathbf{s}_d.
	\end{equation}
	
	This expression encompasses a wide range of practical modulation schemes, such as SC, OFDM, orthogonal time-frequency space (OTFS), and AFDM. For instance, when $\mathbf{U}$ is chosen as the discrete Fourier transform (DFT) matrix, the formulation reduces to OFDM \cite{liu2025uncovering}. More generally, alternative unitary transforms may be employed to realize customized waveform designs in ISAC systems, enabling flexible trade-offs between communication reliability and sensing performance.  
	
	\begin{assumption}[Unit Power and Circularly-Symmetric.]
		\label{Asmp1}
		The transmitted data symbols $s_{d,i}$, for $i = 1, 2, \ldots, N$, are assumed to be independent and identically distributed (i.i.d.) random variables drawn from a standard complex constellation. Specifically, each symbol satisfies the following statistical properties:
		\begin{itemize}
			\item \textbf{Unit variance}: the transmitted symbols are normalized such that $\mathbb{E}\left(|s_{d,i}|^2\right) = 1$;
			\item \textbf{Zero mean}: the symbols are centered around the origin, i.e., $\mathbb{E}\left(s_{d,i}\right) = 0$;
			\item \textbf{Zero pseudo-variance}: $\mathbb{E}\left(s_{d,i}^2\right) = 0$, which implies that the symbols are circularly-symmetric complex random variables.
		\end{itemize}
		Such conditions are commonly adopted in communication systems, as they model practical constellations such as QPSK and QAM \cite{liu2025cp}. 
	\end{assumption}

	\subsection{Received Signal}
	In this paper, we adopt the additive white Gaussian noise (AWGN) channel as a baseline scenario. In practice, sensing performance can be enhanced by \emph{coherent integration}. Specifically, when the target remains stationary over $M$ transmission slots, the transmitter can generate $M$ independent symbol sequences $\{\mathbf{x}_m\}_{m=1}^{M}$ from a chosen constellation. After CP removal, the received signal in the $m$-th slot ($m = 1, 2,\cdots,M$) is then given by
	\begin{equation}
		\mathbf{y}_m =  \beta_0 \mathbf{J}_{n_0} \mathbf{x}_m + \underbrace{\sum_{q=1}^Q \beta_q \mathbf{J}_{n_q} \mathbf{x}_m}_{\mathrm{Clutter}} + \mathbf{n}_m, 
	\end{equation}
	where $\mathbf{J}_{k}$ is the $k$th shift matrix in the form of 
	\begin{equation}
		\mathbf{J}_k = \left[\begin{matrix}
			\mathbf{0} &\mathbf{I}_{N-k}\\
			\mathbf{I}_{k} & \mathbf{0} 
		\end{matrix}\right]. 
	\end{equation}
	The corresponding transmit vector is expressed as
	\begin{equation}
		\mathbf{x}_m = \mathbf{x}_{p,m} + \mathbf{U} \mathbf{s}_{d,m} \in \mathbb{C}^{N \times 1}.
	\end{equation}
	The additive white Gaussian noise is given by $\mathbf{n}_m\sim \mathcal{CN}(\mathbf{0}, \sigma_n^2 \mathbf{I}_N)$, and $\beta_q \in \mathbb{C},\;q=0,\cdots,Q,$ denotes the complex channel gain corresponding to the $q$-th target/scatter point, jointly capturing the effects of propagation loss and reflection coefficient. The matrix $\mathbf{J}_{n_q}$ denotes a circular shift operator that introduces a delay of $n_q$ samples corresponding to the $q$-th target/scatter point. 
	The primary objective of this work is to suppress the sidelobes induced by strong clutter echo returns. These sidelobes often mask signals from a low-observable target, preventing its detection. Accordingly, this paper focuses on a scenario where the target of interest (TOI) possesses a signal power substantially lower than that of the other dominant echoes.

	\section{Adaptive Matched Filter With Random Signals}

	In this section, we first develop an AMF that operates in the presence of superimposed random signals. Owing to the random nature of the received signal, exact finite‑dimensional performance characterization of the AMF is analytically intractable. To this end, we pursue an asymptotic analysis of the AMF output SCNR by utilizing the random matrix theory.

	\subsection{AMF Design}
	During the $m$th time slot, the AMF adaptively adjusts its weight vector $\mathbf{w}_m$ based on the received data to maximize the output SCNR, i.e.,
	\begin{equation}\label{SCNRm}
		\begin{split}
			\gamma_m = \frac{|\beta_0|^2\mathbf{w}_m^\HH \mathbf{J}_{n_0}\mathbf{x}_m \mathbf{x}_m^\HH \mathbf{J}_{-n_0} \mathbf{w}_m}{\mathbf{w}_m^\HH \mathbf{R}_m \mathbf{w}_m},
		\end{split}
	\end{equation}
	where
	\begin{equation}
		\begin{split}
			\mathbf{R}_m &= \mathbb{E}_{\mathbf{n}}\left( {\sum_{q=1}^Q \beta_q \mathbf{J}_{n_q} \mathbf{x}_m} + \mathbf{n}_m \right)\left( {\sum_{q=1}^Q \beta_q \mathbf{J}_{n_q} \mathbf{x}_m} + \mathbf{n}_m \right)^\HH\\
			& = \left( {\sum_{q=1}^Q \beta_q \mathbf{J}_{n_q} \mathbf{x}_m} \right)\left( {\sum_{q=1}^Q \beta_q \mathbf{J}_{n_q} \mathbf{x}_m} \right)^\HH +\sigma_n^2 \mathbf{I}.
		\end{split}
	\end{equation}
	
	The AMF design is equivalent to finding the weight vector $\mathbf{w}_m$ that satisfies 
	\begin{equation}
		\begin{split}\label{Eq00}
			\min_{\mathbf{w}_m}\; &\mathbf{w}_m^\HH \mathbf{R}_m \mathbf{w}_m\\
			s.t. \; & \mathbf{w}_m^\HH\mathbf{J}_{n_0}\mathbf{x}    = 1.
		\end{split}
	\end{equation} 
	This problem is equivalent to the well-known minimum variance distortionless response (MVDR) problem, whose closed-form solution is given by 
	\begin{equation}\label{wopt}
		\mathbf{w}_{m,\dagger}=\frac{1}{ \mathbf{x}_m^\HH\mathbf{J}_{-n_0}\mathbf{R}_m^{-1}\mathbf{J}_{n_0}\mathbf{x}_m} \mathbf{R}_m^{-1}\mathbf{J}_{n_0}\mathbf{x}_m .
	\end{equation}
	
	Substituting \eqref{wopt} into \eqref{SCNRm} yields 
	\begin{equation}\label{SCNRm0}
		\begin{split}
			\gamma_m =|\beta_0|^2 \mathbf{x}_m^\HH\mathbf{J}_{-n_0}\mathbf{R}_m^{-1}\mathbf{J}_{n_0}\mathbf{x}_m.
		\end{split}
	\end{equation}
	
	Note that the matrices $\mathbf{J}_{n_0}$ are unitary (i.e., $\mathbf{J}_n^\HH = \mathbf{J}_n^{-1} = \mathbf{J}_{-n}$) and follow the composition rule $\mathbf{J}_a \mathbf{J}_b = \mathbf{J}_{a+b}$. 
	Define an auxiliary matrix $\mathbf{H}$ as
	\begin{equation}
		\mathbf{H} =\mathbf{J}_{n_0}^\TT \left({\sum_{q=1}^Q \beta_q \mathbf{J}_{n_q } }\right) = {\sum_{q=1}^Q \beta_q \mathbf{J}_{n_q - n_0} }.
	\end{equation}
	Then, we have
	\begin{equation}\label{Qdef}
		\begin{split}
			\mathbf{\Phi}_m&\triangleq\mathbf{J}_{-n_0}\mathbf{R}_m^{-1}\mathbf{J}_{n_0} =\left(\mathbf{H}  \mathbf{x}_m\mathbf{x}_m^\HH\mathbf{H}^\HH +\sigma_n^2 \mathbf{I}\right)^{-1}.
		\end{split} 
	\end{equation}
	Substituting \eqref{Qdef} back into the definition of $\gamma_m$, we arrive at the compact quadratic form:
	\begin{equation}\label{gamma_00}
		\gamma_m =|\beta_0|^2\mathbf{x}_m^\HH\mathbf{\Phi}_m\mathbf{x}_m .
	\end{equation}
	
	\begin{remark}
		The output SCNR is an important metric for evaluating the performance of conventional radar systems. However, a fundamental limitation becomes apparent when this metric is directly applied to ISAC systems. Specifically, the output SCNR is an explicit function of the transmitted waveform. In the ISAC context, the transmitted waveform is modeled by a random data stream, rather than being a pre-designed, deterministic radar pulse. Consequently, the output SCNR itself becomes a random variable, which would fluctuate from one time slot to the next as the data payloads change.
		
		It would pose severe practical challenges to adopt this instantaneous SCNR as the core objective for system design. For instance, a pilot sequence and receiver filter designed to maximize the instantaneous SCNR would be optimal only for the specific data realization within a single time slot. This implies that the system would need to recompute the optimal pilot and reconfigure the receiver for every new transmission slot. Such a continuously adaptive scheme would not only impose a prohibitive computational burden on the real-time processor but also require substantial signaling overhead to support the rapid coordination between the transmitter and receiver.
		
		To address these issues, we propose shifting the optimization objective from an instantaneous figure of merit to a statistical one. Therefore, in the following, we will focus on optimizing the average SCNR over all possible payloads. This approach yields a stable, singular optimization target that depends only on the statistical properties of the communication signal, not its instantaneous values. It thereby enables the design of a fixed pilot scheme and receiver structure that are statistically optimal over the processing block. 
	\end{remark}

	\subsection{Average SCNR}
	
	In this section, we derive a closed-form expression for the average SCNR, i.e.,
	\begin{equation}
		\overline{\gamma}\triangleq \lim\limits_{M\to \infty} \sum_{m = 1}^{M} \gamma_m.
	\end{equation}
	To achieve analytical tractability, we employ an asymptotic approximation based on large-dimensional random matrix theory. This approach allows us to characterize the average system performance in the limit of large signal dimensions, where stochastic fluctuations vanish, and deterministic equivalents can be obtained. The main result is formally stated in the following proposition.

	\begin{proposition}\label{Prop_gammabar}
		Define $\mathbf{\Omega} = \mathbf{x}_p\mathbf{x}_p^\HH$. Let the singular value decomposition (SVD) of $\mathbf{H}$ be given by $\mathbf{H} = \mathbf{U} \mathbf{\Lambda} \mathbf{V}^\HH$, where $\mathbf{\Lambda} = \diag\left(\lambda_1,\cdots,\lambda_N\right)$. Then, we define the following fixed-point equations, i.e., 
		\begin{equation}\label{Tfp_eq}
			\begin{split}
				\left\{\begin{array}{l}
					\mathbf{T} = \left(\mathbf{\Psi}^{-1} + \sigma_n^2 \tilde{\psi} \mathbf{\Lambda}\mathbf{\Omega}\mathbf{\Lambda}^\HH\right)^{-1},\\
					\tilde{t} = \left(\tilde{\psi}^{-1} + \sigma_n^2\tr\left(\mathbf{\Lambda}^\HH \mathbf{\Psi}\mathbf{\Lambda}\mathbf{\Omega}\right)\right)^{-1},\\
				\end{array}
				\right.
			\end{split}
		\end{equation}
		where
		\begin{equation}
			\begin{split}
				\mathbf{\Psi} = \left[\sigma_n^2\left(\mathbf{I} + \tilde{t}\mathbf{\Lambda}\mathbf{\Lambda}^\HH\right)\right]^{-1}, \;
				\tilde{\psi} = \left[\sigma_n^2\left( 1+\tr\left(\mathbf{\Lambda}\mathbf{\Lambda}^\HH \mathbf{T}\right)\right)\right]^{-1}.
			\end{split}
		\end{equation}
		As $N \to \infty$, we have
		\begin{equation}\label{overlinegamma}
			\begin{split}
				\overline{\gamma} \overset{a.s.}{\to}  \frac{|\beta_0|^2}{\sigma_n^2}\left(N + \tr(\mathbf{\Omega}) - \sigma_n^{-2}\eta \left(1-  N +\sigma_n^2 \tr (\mathbf{T})\right)\right) ,
			\end{split}
		\end{equation}
		where the deterministic matrix $\mathbf{T}$ can be obtained by solving \eqref{Tfp_eq}. The deterministic coefficient $\eta$ is given by
		\begin{equation}
			\begin{split}
				\eta =  &\left\vert \tr \left(\mathbf{H} \mathbf{\Omega}\right) \right\vert^2 +\tr \left(\mathbf{H}\mathbf{H}^\HH  \mathbf{\Omega}\right) +\tr \left(\mathbf{H}^\HH \mathbf{H} \mathbf{\Omega}\right)\\
				&+\sum_{q = 1}^{Q}\left(N + (\kappa - 2)\left\Vert \mathbf{b}_{n_q - n_0}\right\Vert_2^2 \right),
			\end{split}
		\end{equation}
		where $\kappa$ denotes the kurtosis of the constellation, and 
		\begin{equation}
			\begin{split}
				\mathbf{b}_k = \left[\sum_{n=1}^{N} |v_{1,n}|^2 e^{\frac{-i2\pi k (n-1)}{N}},\cdots,\sum_{n=1}^{N} |v_{N,n}|^2 e^{\frac{-i2\pi k (n-1)}{N}}\right]^\TT,
			\end{split}
		\end{equation}
		with $v_{m,n}$ being the $(m,n)$th entry of $\mathbf{V} = \mathbf{U}^\HH \mathbf{F}_N^\HH$.

	\end{proposition}
	
	\textbf{\emph{Proof}}: See Appendix \ref{Proof_Prop_gammabar}.\hfill $\blacksquare$
	
	To provide clearer physical intuition, we state the following corollary, which establishes an explicit upper bound on the average SCNR.
	
	\begin{corollary}\label{CorUBLB}
		The upper bound of the average SCNR is given by
		\begin{equation}
			\begin{split}
				\overline{\gamma}_{\mathrm{UB}}  =  \frac{|\beta_0|^2 (N + \tr(\mathbf{\Omega}))}{\sigma_n^2}.
			\end{split}
		\end{equation}
	\end{corollary}
	
	\textbf{\emph{Proof}}: 
	From \eqref{gammadef1}, we have
	\begin{equation}
		\begin{split}
			\overline{\gamma} \leq \frac{|\beta_0|^2 }{\sigma_n^{2} }\mathbb{E}\left(\mathbf{x}^\HH \mathbf{x}\right) =  \frac{|\beta_0|^2 (N + \tr(\mathbf{\Omega}))}{\sigma_n^2} \triangleq \overline{\mathrm{SNR}}.
		\end{split}
	\end{equation}
	The equality holds when the clutter power $\beta_q, q=1,2,\cdots, Q$ is much smaller than the noise power. In this case, the SCNR will approach the SNR.
	\hfill $\blacksquare$

	\begin{remark}
		From \textbf{\emph{Corollary \ref{CorUBLB}}} and its proof, we can observe that, the upper bound can be achieved in the ideal case where the total power of all clutter signals is much smaller than the power of the background noise (i.e., $|\beta_q|^2 \ll \sigma_n^2, q=1,2,\cdots,Q$). In this case, the impact of the clutter term in the denominator of SCNR becomes negligible. This situation typically occurs in what are known as “noise-dominant” communication environments.
	\end{remark}
    
		\begin{corollary}\label{MONO}
			The average SCNR $\overline{\gamma}$ decreases monotonically with the kurtosis $\kappa$ in the asymptotic regime. 
		\end{corollary}
		
		\textbf{\emph{Proof}}: See Appendix \ref{proofMONO}. \hfill $\blacksquare$

		\begin{remark}\label{Cons_Modu}
		According to \cite[Table II]{liu2025cp}, the kurtosis $\kappa$ for PSK, QAM, (e.g. 16-QAM), and Gaussian symbols  are $1$, $1.32$, and $2$, respectively. 
		This corollary indicates that, for a given modulation basis, PSK achieves the highest average SCNR among these constellations. The superior performance of the PSK comes from its constant modulus property. All symbols in 4-PSK and 16-PSK have the same power envelope, creating a stable waveform structure. This uniformity is highly beneficial for radar sensing, as it simplifies tasks like clutter suppression and target detection. In contrast, 16-QAM is a non-constant modulus constellation, whose symbols have varying power levels. This amplitude fluctuation introduces an additional degree of randomness into the transmitted waveform, thereby slightly degrading the SCNR. 
		The Gaussian payload represents the extreme case, highlighting a fundamental trade-off in ISAC design. While a Gaussian input is known to achieve the Shannon capacity limit and is thus optimal for pure communication, its complete lack of deterministic structure makes it the least effective for sensing. From a radar perspective, its random, noise-like nature makes it extremely difficult to distinguish desired target echoes from actual clutter and noise, which leads to the observed significant performance degradation.
		
		Furthermore, \textbf{\emph{Corollary \ref{MONO}}} indicates that our findings corroborate those of \cite{liu2025cp}: OFDM attains the highest average SCNR. This advantage is attributable to OFDM yielding the smallest value of $\eta$, which quantifies the effective sidelobe level. Consequently, this means that OFDM is the most effective at mitigating clutter in the considered scenarios. 
		\end{remark}

	\section{Pilot design: DPD and DPI schemes}
	In the following, we present the details of the proposed DPD and DPI pilot design schemes, respectively. Since $\beta_0$ is a constant which independent of pilots, we omit it in the following optimization problem for simplification. 
    
	\subsection{DPD Pilot design}
	In the $m$th time slot, given a realization of $\mathbf{s}_{d,m}$, the DPD pilot design scheme requires that the pilot sequence $\mathbf{x}_{p,m}$ should be adaptively optimized according to  $\mathbf{s}_{d,m}$.  
	For notational simplicity, the subscript $m$ is omitted hereafter, and thus, $\mathbf{x}_{p,m}$ and $\mathbf{s}_{d,m}$ are denoted by $\mathbf{x}_{p}$ and $\mathbf{s}_{d}$, respectively. 
	The DPD pilot design problem can be formulated by
	\begin{equation} \label{Prob_DPD_0}
		\begin{split}
			\max_{\mathbf{x}_{p}} \; & \mathcal{F}(\mathbf{x}_{p})\\
			s.t. \; & \Vert\mathbf{x}_{p}\Vert^2 \leq P_p N, 
		\end{split}
	\end{equation}
	where $P_p$ denotes the power of pilot and 
	\begin{equation} 
		\begin{split}
			\mathcal{F}(\mathbf{x}_{p}) \triangleq \gamma = \mathbf{x}^\HH\mathbf{\Phi}^{-1}\mathbf{x}.
		\end{split}
	\end{equation}

	Note that problem \eqref{Prob_DPD_0} takes the form of a multidimensional single‑ratio fractional program \cite{shen2018fractional,xie2022perceptive}. By utilizing Dinkelbach’s transform, it is equivalent to a sequence of non‑fractional problems, i.e., 
	\begin{equation} \label{Prob_DPD_1}
		\begin{split}
			\max_{\mathbf{x}_{p},\mathbf{u}} \; & 
			2 \Re\left(\mathbf{u}^\HH \mathbf{x}\right) - \mathbf{u}^\HH \left(\mathbf{H}  \mathbf{x}\mathbf{x}^\HH\mathbf{H}^\HH +\sigma_n^2 \mathbf{I}\right)\mathbf{u}\\
			s.t. \; & \Vert\mathbf{x}_{p}\Vert^2 \leq P_p N.
		\end{split}
	\end{equation}
	
	To solve this problem, an iteration process based on AO is introduced. In particular, given a feasible initial point $\left\{\mathbf{x}_{p,(0)},\mathbf{u}_{(0)}\right\}$, at the $t$th iteration, we iteratively 
	\begin{itemize}
		\item update $\mathbf{u}_{(t+1)}$ with fixed $\mathbf{x}_{p,(t)}$,
		\item update $\mathbf{x}_{p,(t+1)}$ with fixed $\mathbf{u}_{(t+1)}$.
	\end{itemize}
	These steps are repeated until convergence, which produces a stationary point of the problem \eqref{Prob_DPD_1}.
	
	\subsubsection{Update $\mathbf{u}_{(t+1)}$ with fixed $\mathbf{x}_{p,(t)}$}
	The closed-form solution of $\mathbf{u}_{(t+1)}$ is given by
	\begin{equation}
		\label{ut1}
		\begin{split}
			\mathbf{u}_{(t+1)} = \left(\mathbf{H}  \mathbf{x}_{(t)}\mathbf{x}_{(t)}^\HH\mathbf{H}^\HH +\sigma_n^2 \mathbf{I}\right)^{-1}\mathbf{x}_{(t)},
		\end{split}
	\end{equation}
	where
	\begin{equation}
		\begin{split}
			\mathbf{x}_{(t)} = \mathbf{x}_{p,(t)} + \mathbf{U}\mathbf{s}_d.
		\end{split}
	\end{equation}

	\subsubsection{Update $\mathbf{x}_{p,(t+1)}$ with fixed $\mathbf{u}_{(t+1)}$}
	The objective function at the $t$th iteration can be reformulated by
	\begin{equation}\label{Ft1}
		\mathcal{F}\left(\mathbf{x}_p | \mathbf{u}_{(t+1)}\right) = 2 \Re \left(\mathbf{c}_{(t+1)}^\HH \mathbf{x}_p\right) -\mathbf{x}_p^\HH \mathbf{B}_{(t+1)} \mathbf{x}_p + C_{(t+1)},
	\end{equation}
	where
	\begin{align}
		\mathbf{c}_{(t+1)} &= \mathbf{u}_{(t+1)} - \mathbf{H}^\HH \mathbf{u}_{(t+1)}\mathbf{u}_{(t+1)}^\HH\mathbf{H} \mathbf{U} \mathbf{s}_d,\\
		\mathbf{B}_{(t+1)} &= 
		\mathbf{H}^\HH \mathbf{u}_{(t+1)}\mathbf{u}_{(t+1)}^\HH\mathbf{H},\\
		\begin{split}
			C_{(t+1)} &= -  \mathbf{s}_d^\HH\mathbf{U}^\HH\mathbf{H}^\HH \mathbf{u}_{(t+1)}\mathbf{u}_{(t+1)}^\HH\mathbf{H} \mathbf{U} \mathbf{s}_d  \\
			&\quad - \mathbf{u}_{(t+1)}^\HH \mathbf{u}_{(t+1)}+ 2\Re\left(\mathbf{u}_{(t+1)}^\HH\mathbf{U}\mathbf{s}_d\right).
		\end{split} 
	\end{align} 
	Then the problem can be reformulated as the maximization of \eqref{Ft1}, which can be efficiently solved by the Lagrange multiplier method. Specifically, we introduce a penalty function to reformulate the problem in \eqref{Ft1} as an unconstrained optimization problem that minimizes
	\begin{equation}\label{Ft12}
		\mathcal{L}\left(\mathbf{x}_p\right) = -\mathcal{F}\left(\mathbf{x}_p | \mathbf{u}_{(t+1)}\right) + \gamma_p \left(\mathbf{x}_p^\HH \mathbf{x}_p - 1\right),
	\end{equation} 
	where $\gamma_p$ is the Lagrange penalty coefficient. Note that \eqref{Ft12} is convex w.r.t. $\mathbf{x}_p$. The minimizer of \eqref{Ft12} can be obtained by solving $\nabla_{\mathbf{x}_p} \mathcal{L}\left(\mathbf{x}_p\right) = \mathbf{0}$, i.e.,
	\begin{equation}\label{xpdagger}
		\mathbf{x}_{p,\mathrm{opt}} = \left(\mathbf{B}_{(t+1)} + \gamma_p \mathbf{I}\right)^{-1} \mathbf{c}_{(t+1)}.
	\end{equation} 
	Since $\mathbf{x}_{p,\mathrm{opt}}$ depends on $\gamma_p$, it is necessary to find a suitable $\gamma_p$. By performing SVD $\mathbf{B}_{(t+1)} = \mathbf{V}_B \mathbf{\Lambda}_B \mathbf{V}_B^\HH$, according to the complementary Karush–Kuhn–Tucker (KKT) condition, we have
	\begin{equation}\label{xp2}
		||\mathbf{x}_{p,\mathrm{opt}}||^2 = \mathbf{x}_{p,\mathrm{opt}}^\HH \mathbf{x}_{p,\mathrm{opt}} = \sum_{i = 1}^{N} \frac{|\mathbf{v}_{B,i}^\HH \mathbf{c}_{(t+1)}|}{\left(\lambda_{B,i} + \gamma_p\right)^2},
	\end{equation} 
	where $\mathbf{v}_{B,i}$ denotes the $i$th column of $\mathbf{V}_{B}$ and $\lambda_{B,i}$ is the $(i,i)$th entry of $\mathbf{\Lambda}_{B}$. It is easy to check that $\mathbf{x}_{p,\mathrm{opt}}^\HH$ is monotonic w.r.t. $\gamma_p$. We thus utilize the bisection method to find a suitable $\gamma_p$ to make $||\mathbf{x}_{p,\mathrm{opt}}||^2 = P_p N$. We then update $\mathbf{x}_{p,(t+1)} = \mathbf{x}_{p,\mathrm{opt}}$. The proposed DPD algorithm is summarized in \textbf{\emph{Alg. \ref{alg1}}}.
	
	\begin{algorithm}[!t]
		\caption{The proposed DPD-based method for solving (\ref{Prob_DPD_0})}
		\begin{algorithmic}[1]
			\STATE Initialize $\mathbf{u}_{(0)}$ and $\mathbf{x}_{p,(0)}$.
			\STATE \textbf{Repeat} 
			\STATE \hspace{0.5cm} Update $\mathbf{u}_{t+1}$ via \eqref{ut1}.
			\STATE \hspace{0.5cm} Compute $\gamma_p$ via the bisection method to make $||\mathbf{x}_{p,\mathrm{opt}}||^2$ in \eqref{xp2} equals to $P_p N$.
			\STATE \hspace{0.5cm} Update $\mathbf{x}_{p,(t+1)} = \mathbf{x}_{p,\mathrm{opt}}$ via \eqref{xpdagger}. 
			\STATE \hspace{0.5cm} $t \gets t+1$.
			\STATE \textbf{Until} Convergence criterion is met. 
			\STATE \textbf{return}  $\mathbf{x}_{p,\star} =\mathbf{x}_{p,(t)}$.
		\end{algorithmic}
		\label{alg1}
	\end{algorithm}

	\begin{remark}
		According to \textbf{\emph{Corollary \ref{CorUBLB}}}, the upper bound of the instantaneous SCNR coincides with the instantaneous SNR. Therefore, given $\mathbf{s}_d$, the performance upper bound of the DPD scheme can be obtained by solving the following problem, i.e.,
		\begin{equation}\label{ProbISNR}
			\begin{split}
				\max_{\mathbf{x}_p} &\left\Vert \mathbf{x}_p +\mathbf{x}_d \right\Vert^2\\
				s.t. & \left\Vert \mathbf{x}_p \right\Vert^2 = P_p N.
			\end{split}
		\end{equation}
		The solution to \eqref{ProbISNR} is given by 
		\begin{equation}
			\mathbf{x}_{p,\mathrm{UB}} = \frac{\sqrt{P_p N}}{\left\Vert \mathbf{x}_d\right\Vert}\mathbf{x}_d.
		\end{equation}
		Then, the performance upper bound of DPD scheme is given by 
		\begin{equation}\label{DPDub}
			\begin{split}
				\gamma_{\mathrm{UB}} = \frac{|\beta_0|^2}{\sigma_n^2}||\mathbf{x}_{p,\mathrm{UB}} +\mathbf{x}_d||^2 = \frac{|\beta_0|^2}{\sigma_n^2}\left(1+\frac{\sqrt{P_pN}}{||\mathbf{x}_d||}\right)^2 ||\mathbf{x}_d||^2.
			\end{split}
		\end{equation}
This upper bound provides valuable insight into the theoretical performance limit of the system. It characterizes the maximum achievable signal quality under ideal (clutter-free) conditions and serves as a benchmark against which practical algorithms and pilot designs can be evaluated. By comparing the actual SCNR with this upper bound, we can quantify the efficiency of the proposed scheme and identify how closely it approaches the fundamental limit.

It is observed that the upper bound for the DPD scheme exceeds that of the DPI scheme. This is because the constant DPI pilot is assumed independent of the data payload, which yields $\mathbb{E}(|\mathbf{x}_p^\HH \mathbf{x}_d|^2) = 0$. In contrast, under DPD the pilot $\mathbf{x}_p$ is designed for each realization of the data payload $\mathbf{x}_d$, and hence generally depends on $\mathbf{x}_d$. Consequently, the cross term $|\mathbf{x}_p^\HH \mathbf{x}_d|^2$ has a nonzero expectation in general, and the resulting upper bound of SNR differs and can be larger. 
	\end{remark}
	
	\subsection{DPI Pilot design}
	
	In the DPI scheme, the pilots remain unchanged over $M$ slots. 
	The DPI scheme can be formulated by
	\begin{equation} \label{Prob_DPI_0}
		\begin{split}
			\max_{\mathbf{\Omega}} \; & \overline{\gamma}\\
			s.t. \; & \tr(\mathbf{\Omega}) \leq P_p N, \quad  \mathrm{Rank}(\mathbf{\Omega}) = 1,
		\end{split}
	\end{equation}
	where $\mathbf{\Omega} \triangleq \mathbf{x}_p\mathbf{x}_p^\HH$ and $\overline{\gamma}$ denotes the average SCNR, i.e.,
	\begin{equation} 
		\begin{split}
			\overline{\gamma}  = \lim\limits_{M \to \infty} \frac{1}{M} \sum_{m=1}^{M}\gamma_m = \mathbb{E}_{\mathbf{s} }  \mathcal{F}(\mathbf{s} ) .
		\end{split}
	\end{equation}

	Since the objective function in problem \eqref{Prob_DPI_0} is defined implicitly through a fixed-point equation, the problem is inherently challenging to solve. In particular, an explicit expression for the Hessian matrix is challenging to obtain, as the matrix $\mathbf{T}$ itself is determined by solving a fixed-point equation that depends on $\mathbf{x}_p$. Moreover, the constraint lies on a fixed-rank manifold, which further complicates the optimization process. To address these difficulties, we propose to solve problem \eqref{Prob_DPI_0} by leveraging manifold optimization and the gradient-projection (GP) method. This approach naturally accommodates the manifold constraint and relies solely on gradient information, thereby avoiding the need for higher-order derivatives.

	To derive the gradient of $\overline{\gamma}$ with respect to $\mathbf{\Omega}$, we must first establish the gradients of its constituent components, namely  $\eta$ and $\tr \mathbf{T}$.
	
	1) The gradients of $\eta$ with respect to $\mathbf{\Omega}$, i.e., $\mathbf{\Delta}_\eta \triangleq \frac{\partial\eta}{\partial \mathbf{\Omega}}$, is given by
	\begin{equation}\label{Deltaeta}
		\begin{split}
			\mathbf{\Delta}_\eta  = \tr\left(\mathbf{H}\mathbf{\Omega}\right) \mathbf{H}^\HH + \tr\left(\mathbf{H}^\HH\mathbf{\Omega}\right) \mathbf{H}
			+  \mathbf{H}\mathbf{H}^\HH  +\mathbf{H}^\HH\mathbf{H}.
		\end{split}
	\end{equation}
	
	2) The gradients of $\tr \mathbf{T}$ with respect to $\mathbf{\Omega}$ is denoted by $\mathbf{\Delta}_{T}\triangleq \frac{\partial \tr \mathbf{T}}{\partial \mathbf{\Omega}}$, whose $(i,j)$th entry is given by
	\begin{equation}\label{Deltaij}
		\begin{split}
			\left[\mathbf{\Delta}_{T} \right]_{i,j}= \tr \left(\mathbf{T}_{i,j}'\right).
		\end{split}
	\end{equation}
	Here, $\mathbf{T}_{i,j}'$ is obtained  by solving the fixed-point equations:	
	\begin{equation}\nonumber 
		\begin{split}
			\left\{\begin{array}{l}
				\mathbf{T}_{i,j}'  =-\mathbf{T}\left( \underline{\mathbf{\Psi}}_{i,j}'
				+ \sigma_n^2\tilde{\psi}_{i,j}' \mathbf{\Lambda}\mathbf{\Omega}\mathbf{\Lambda}^\HH+\sigma_n^2 \tilde{\psi} \mathbf{\Lambda}\mathbf{e}_j\mathbf{e}_i^\TT\mathbf{\Lambda}^\HH
				\right)\mathbf{T},\\
				\tilde{t}_{i,j}' =-{\tilde{t}^2}{\left(\underline{\tilde{\psi}}_{i,j}' 
					+ \sigma_n^2 \mathbf{e}_i^\TT\mathbf{\Lambda}^\HH \mathbf{\Psi}\mathbf{\Lambda}\mathbf{e}_j+\sigma_n^2 \mathbf{x}_p^\HH\mathbf{\Lambda}^\HH \mathbf{\Psi}_{i,j}' \mathbf{\Lambda} \mathbf{x}_p
					\right)},
			\end{array}
			\right.
		\end{split}
	\end{equation}
	where $\mathbf{e}_i$ denotes the vector whose $i$th entry is $1$ and all other entries are $0$, and
	\begin{equation}
		\begin{split}
			\underline{\mathbf{\Psi}}_{i,j}'&\triangleq\frac{\partial \mathbf{\Psi}^{-1}}{\partial \Omega_{i,j}^*} = \sigma_n^2\tilde{t}_{i,j}' \mathbf{\Lambda}\mathbf{\Lambda}^\HH,\\
			\mathbf{\Psi}_{i,j}'&\triangleq\frac{\partial \mathbf{\Psi}}{\partial \Omega_{i,j}^*} = -\sigma_n^2  \tilde{t}_{i,j}' \mathbf{\Psi} \mathbf{\Lambda}\mathbf{\Lambda}^\HH \mathbf{\Psi}= -\mathbf{\Psi} \underline{\mathbf{\Psi}}_{i,j}'\mathbf{\Psi},\\
			\underline{\tilde{\psi}}_{i,j}' &\triangleq \frac{\partial \tilde{\psi}^{-1}}{\partial \Omega_{i,j}^*} = \sigma_n^2 \tr \left(\mathbf{\Lambda}\mathbf{\Lambda}^\HH \mathbf{T}_{i,j}'\right),\\
			\tilde{\psi}_{i,j}'&\triangleq\frac{\partial \tilde{\psi}}{\partial \Omega_{i,j}^*} = -\frac{\sigma_n^2\tr \left(\mathbf{\Lambda}\mathbf{\Lambda}^\HH \mathbf{T}_{i,j}'\right)}{\left[\sigma_n^2 \left(1+\tr \left(\mathbf{\Lambda}\mathbf{\Lambda}^\HH\mathbf{T}\right)\right)\right]^2} =  - \underline{\tilde{\psi}}_{i,j}'\tilde{\psi}^2.
		\end{split}
	\end{equation}
	
	Then, the gradient of $\overline{\gamma}$ with respect to $\mathbf{\Omega}$ is given by
	\begin{equation}\label{deltat}
		\begin{split}
			&\mathbf{\Delta}_{(t)} \triangleq \mathbf{\nabla}_{\mathbf{\Omega}}\overline{\gamma}|_{\mathbf{\Omega}=\mathbf{\Omega}_{(t)}}\\
			& = \frac{|\beta_0|^2}{\sigma_n^2}\left[\mathbf{I} - \sigma_n^{-2} \left(1-N + \sigma_n^2 \tr \mathbf{T} \right) \mathbf{\Delta}_{\eta,(t)} - \eta \mathbf{\Delta}_{T,(t)}\right],
		\end{split}
	\end{equation}
	where $\mathbf{\Delta}_{\eta,(t)} = \mathbf{\Delta}_\eta|_{\mathbf{\Omega}=\mathbf{\Omega}_{(t)}}$ and $\mathbf{\Delta}_{T,(t)} = \mathbf{\Delta}_{T}|_{\mathbf{\Omega}=\mathbf{\Omega}_{(t)}}.$

	The manifold-based method updates the variable within its tangent space. By updating along the tangent space with a small enough step, the new point is almost within the feasible set. For the fixed-rank manifold, the projection operator onto the tangent space is
	\begin{equation}
		\mathcal{P} \left(\mathbf{X}\right)= \mathbf{P}_U \mathbf{X} \mathbf{P}_U + \mathbf{P}_U^\perp \mathbf{X} \mathbf{P}_V +
		\mathbf{P}_U \mathbf{X} \mathbf{P}_V^\perp,
	\end{equation}
	where $\mathbf{P}_U = \mathbf{u}_{\max}\mathbf{u}_{\max}^\HH$,  $\mathbf{P}_V = \mathbf{v}_{\max}\mathbf{v}_{\max}^\HH$, $\mathbf{P}_U^\perp = \mathbf{I} - \mathbf{P}_U$, and $\mathbf{P}_V^\perp = \mathbf{I} - \mathbf{P}_V$, with 
	$\mathbf{u}_{\max}$ and $\mathbf{v}_{\max}$ denoting the main left and right singular vector of $\mathbf{X}$. Thus, the Riemannian gradient at the $t$th iteration is given by 
	\begin{equation}\label{RieGra}
		\mathbf{G}_{(t)} = \mathrm{grad} \left(\overline{\gamma}\right) = \mathcal{P}\left(\mathbf{\Delta}_{(t)}\right).
	\end{equation}
	Then, $\mathbf{\Omega}$ is updated by
	\begin{equation}\label{update_Omega}
		\widetilde{\mathbf{\Omega}}_{(t+1)} = \mathbf{\Omega}_{(t)} - \alpha_{(t)} \mathbf{G}_{(t)}, 
	\end{equation}
	where $\alpha_{(t)}$ denotes the step size obtained via the Armijo line search. A retraction is needed to remap the updated points onto the feasible region, which is defined as
	\begin{equation}\label{retraction1}
		\begin{split}
			\mathbf{\Omega}_{(t+1)} = \left\{
			\begin{matrix}
				\widetilde{\mathbf{\Omega}}_{(t+1)}, & \mathrm{if}\;  \tr(\mathbf{\Omega}) \leq P_p N\\
				\frac{P_p N}{\tr\left(\widetilde{\mathbf{\Omega}}_{(t+1)}\right)}\widetilde{\mathbf{\Omega}}_{(t+1)}, &\mathrm{otherwise}
			\end{matrix}
			\right. .
		\end{split}
	\end{equation}
	
	The proposed DPI algorithm is summarized in \textbf{\emph{Algorithm \ref{alg2}}}.
	The convergence of the proposed method is guaranteed by \cite[Theorem 4.3.1]{absil2008optimization}

	\begin{algorithm}[!t]
		\caption{The proposed DPI-based method for solving (\ref{Prob_DPI_0})}
		\begin{algorithmic}[1]
			\STATE Initialize $\mathbf{\Omega}_{(0)}$.
			\STATE \textbf{Repeat}
			\STATE \hspace{0.5cm}Compute the gradient $\mathbf{\Delta}_{(t)}$ via \eqref{deltat}. 
			\STATE \hspace{0.5cm}Compute the Riemannian gradient $\mathbf{G}_{(t)}$ via  \eqref{RieGra}.
			\STATE \hspace{0.5cm}Compute $\alpha_{(t)}$ via the Armijo line search step.
			\STATE \hspace{0.5cm}Update $\widetilde{\mathbf{\Omega}}_{(t+1)}$ via \eqref{update_Omega}.
			\STATE \hspace{0.5cm}Update $\mathbf{\Omega}_{(t+1)}$ via the retraction defined in \eqref{retraction1}.
			\STATE \hspace{0.5cm} $t \gets t+1$.
			\STATE \textbf{Until} Convergence criterion is met. 
			\STATE \textbf{return} $\mathbf{\Omega}_{\star}=\mathbf{\Omega}_{(t)}$.
		\end{algorithmic}
		\label{alg2}
	\end{algorithm}

	\section{Simulation Results}
	
	In this section, we validate the accuracy of the theoretical analysis and evaluate the performance of the proposed DPD and DPI pilot design Schemes through simulations. The power of noise is set as $-90$ dBm, respectively. Unless specified otherwise, the number of clutter components is set as $Q = 4$, and the number of snapshots in one time slot is set as $N = 128$. 
	The path gain for the $q$-th clutter path is modeled as $$\beta_q \sim \mathcal{CN}\left(0, 10^{-0.1\vartheta(d_q)}\right),$$
	where $\vartheta(d) = a + 10b\log_{10}(d) + \epsilon$, with $d$ denoting the propagation distance and $\epsilon \sim \mathcal{CN}\left(0, \sigma_{\epsilon}^2\right)$ \cite{6834753}. 
	Following the parameter settings in \cite{6834753}, we use $a = 61.4$, $b = 2$, and $\sigma_{\epsilon} = 5.8$~dB. The distances between the transmitter and the clutter points are randomly generated within the range of $[30,40]$ m. 
	Note that in 5G NR systems, pilot symbols typically occupy only about 10$\%$ of the total resources. In our simulation, the transmit powers of the pilot and data payload are set to $20$ dBm and $30$ dBm, respectively.

	\subsection{Comparison between MF and AMF}
	\begin{figure}[!t]
		\centering
		\includegraphics[width=3.7in]{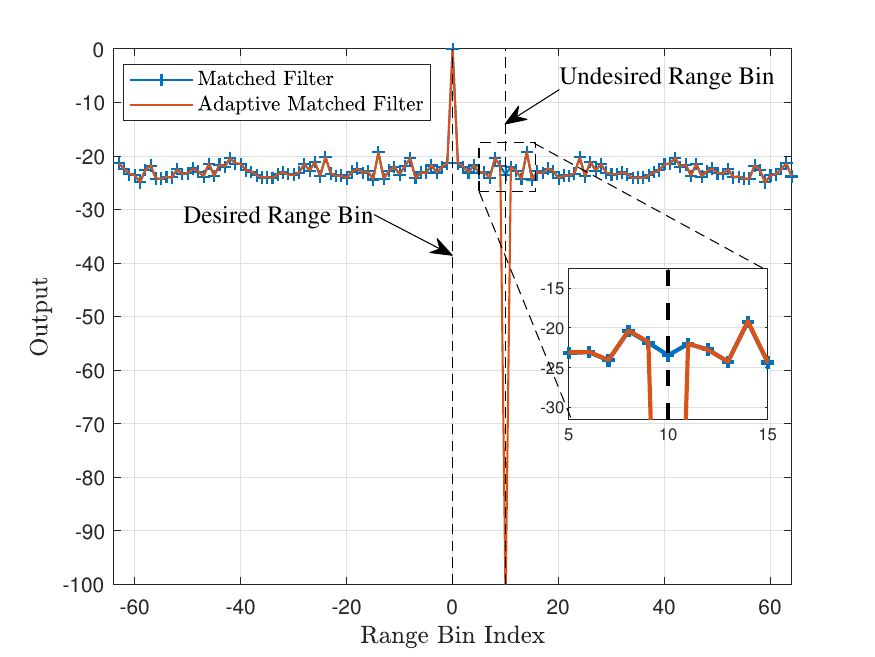}
		\caption{Output of two temporal filters: MF and AMF.}
		\label{ill_AMF}
	\end{figure}

	Fig. \ref{ill_AMF} presents a comparative analysis of the outputs from the MF and the AMF. 
	For this simulation, we set $N = 128$. A single clutter source is located at the $10$th range bin, while the desired target is positioned at the $0$th range bin. It can be observed that, for most range bins, the outputs of the MF and AMF are nearly identical.
	The key difference between the two filters lies in their ability to suppress signals in undesired range bins, particularly at the range bin of clutter. At the $10$th range bin, the MF output exhibits a level of approximately $-23$ dB. In contrast, the AMF, leveraging its adaptive nature, effectively suppresses the output at the same location to about $-100$ dB, achieving nearly $80$ dB of additional clutter rejection. 
	This result indicates that in the presence of strong clutter, sidelobe leakage from the MF can obscure weak target echoes, while the AMF effectively mitigates this issue by forming an adaptive null in the direction of the clutter, thereby significantly improving target detection performance in clutter-dominated environments.

	\subsection{Validation of \textbf{Proposition \ref{Prop_gammabar}}}
	
	\begin{figure}[!t]
		\centering
		\includegraphics[width=3.7in]{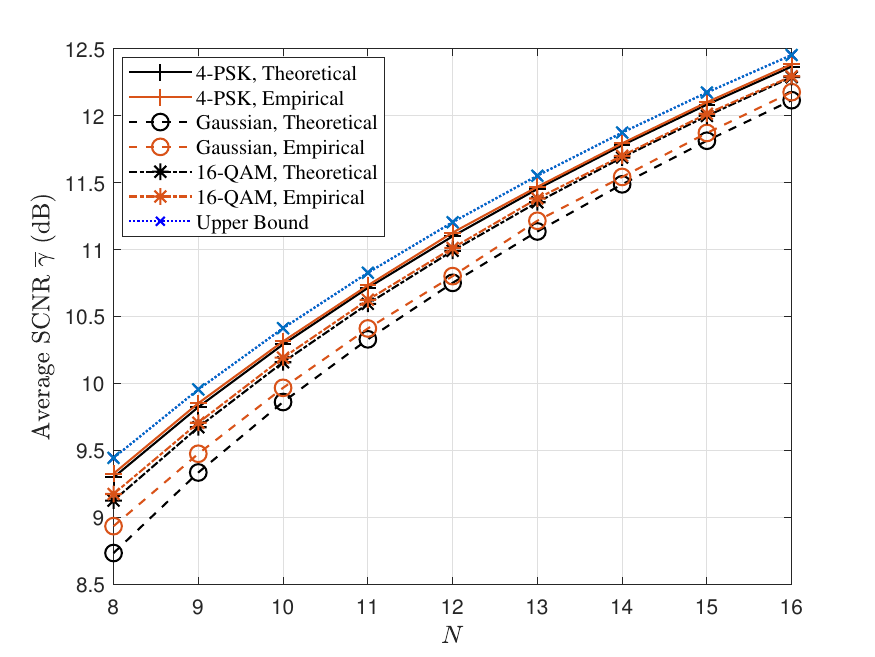}
		\caption{The average SCNR versus the number of snapshots $N$ under different constellations.}
		\label{ill_Vali1}
	\end{figure}

		Fig. \ref{ill_Vali1} illustrates the average SCNR  versus the number of snapshots $N$. 
		We consider the target is located at the $10$th range bin, and one clutter point is located at the $12$th range bin. 
		The legend ``Empirical'' represents the performance obtained through Monte Carlo simulations, where each abscissa is averaged over $10,000$ independent trials.
		The legend ``Theoretical'' represents the deterministic equivalent for the average SCNR, which is defined in \eqref{overlinegamma}. The modulation is selected as OFDM.

		As illustrated in Fig. \ref{ill_Vali1}, we observe the remarkable accuracy of the approximation provided in \textbf{\emph{Proposition \ref{Prop_gammabar}}}, even in non-asymptotic scenarios. Although the proposition is formally derived in the asymptotic regime where $N$ approaches infinity, our numerical results demonstrate its validity for finite dimensions. Specifically, the approximation holds with high precision even if  $N$ is moderately large (e.g., $N\geq 12$). Furthermore, the figure confirms the intuitive trend that the accuracy of the approximation improves as the number of snapshots increases, which causes the empirical results to converge more closely to their deterministic equivalent.
		
		Moreover, we observe that the gap between the simulated $\overline{\gamma}$, and our deterministic equivalent in \eqref{overlinegamma} is consistently smaller than the gap between $\overline{\gamma}$ and the previously established upper bound $\overline{\gamma}_{\mathrm{UB}}$, defined in \textbf{\emph{Corollary \ref{CorUBLB}}}. This demonstrates that our proposed deterministic equivalent provides a more exact approximation, which can serve as a better performance metric for system design.

		\begin{figure}[!t]
			\centering
			\includegraphics[width=3.7in]{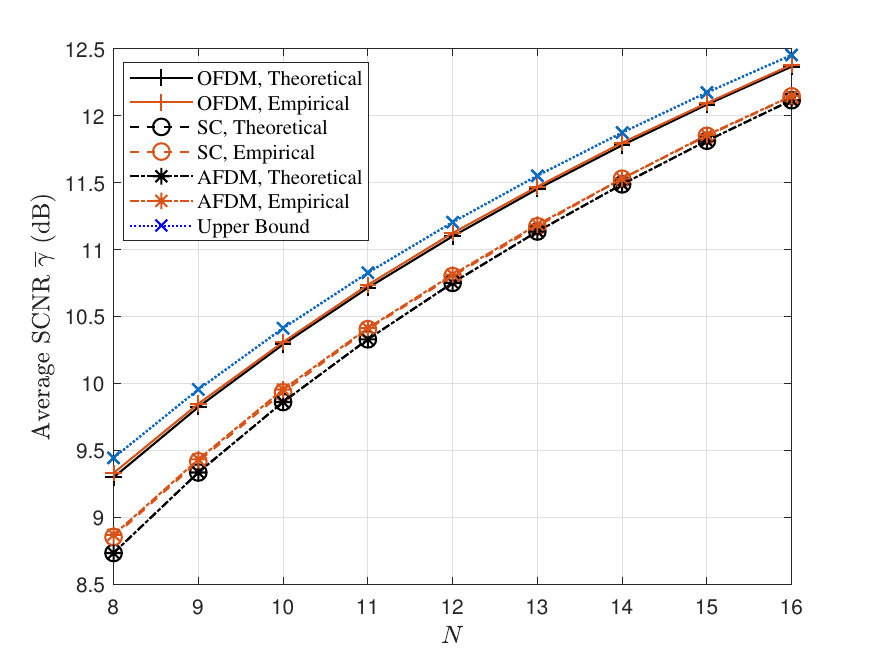}
			\caption{The average SCNR  versus the number of snapshots $N$ under different modulation.}
			\label{ill_Modu}
		\end{figure}

		\subsection{Effect of Constellation and Modulation}
		Fig. \ref{ill_Vali1} also presents the average SCNR when different communication constellations are embedded within the ISAC waveform. 
		In ISAC systems, the data payload is modulated using various schemes. To investigate the impact of this modulation on sensing performance, we evaluated four common choices: 4-PSK, 16-QAM, and a Gaussian distribution. For a fair comparison across these schemes, the all-one pilot was employed as the base sensing sequence for all data payload types, which is widely utilized \cite{wang2020compressed} and serves as a common benchmark in the literature. The results in Fig. \ref{ill_Vali1} reveal a clear performance hierarchy among the constellations, which agrees with \textbf{\emph{Remark \ref{Cons_Modu}}}. 
		First, we observe that 4-PSK achieves the highest SCNR. Following them, 16-QAM exhibits a marginally lower SCNR. The Gaussian-distributed data payload results in the worst sensing performance by a significant margin.
		This performance trend is directly attributable to the structural properties of the constellations, particularly their modulus variation. 
		
		Fig. \ref{ill_Modu} shows the average SCNR with different modulation. For the AFDM, we choose its intermediate coefficient $c_1 = \frac{1}{4N}$ and $c_2 = \frac{1}{2N}$. It is evident that OFDM consistently outperforms both SC and AFDM, which agrees with the results reported in \cite{liu2025cp}. Therefore, unless specified otherwise, we adopt OFDM with 4‑PSK modulation hereafter.

		\subsection{DPD Pilot Design Scheme}
		\begin{figure}[!t]
			\centering
			\includegraphics[width=3.7in]{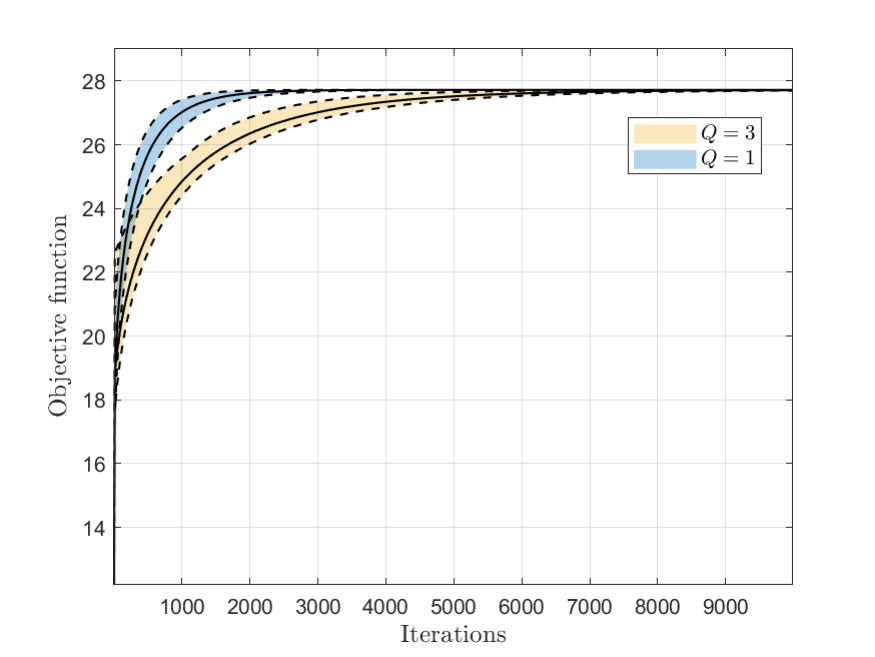}
			\caption{Convergence performance of the proposed DPD pilot design method.}
			\label{ill_DPD_Converge}
		\end{figure}

		\subsubsection{Convergence}
		Fig. \ref{ill_DPD_Converge} illustrates the convergence behavior of the proposed DPD pilot design algorithm. 
		The number of snapshots is set as $N = 16$. There are two clutter points located in the $12$th and $13$th range bins. The target is located at the $10$th range bin. 
		The solid line denotes the average value of the objective function, while the shaded region indicates its fluctuation range across different realizations. It can be observed that the convergence speed is influenced by the number of clutter components, denoted by $Q$. Specifically, as $Q$ increases, the optimization landscape becomes more complex, which may slightly slow down convergence. Nevertheless, as shown in Fig. \ref{ill_DPD_Converge}, the proposed method consistently converges within a few iterations for various values of $Q$, demonstrating its robustness.

		\begin{figure}[!t]
			\centering
			\includegraphics[width=3.7in]{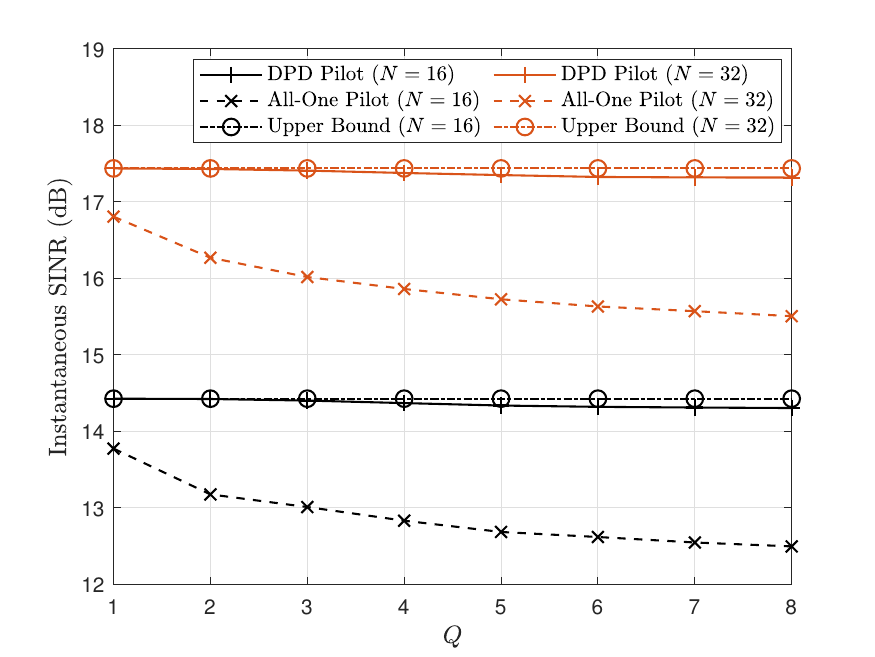}
			\caption{The average SCNR versus the number of clutter components $Q$. }
			\label{DPD_Perform1}
		\end{figure}

		\subsubsection{Performance}
		Fig. \ref{DPD_Perform1} shows the average SCNR versus the number of clutter components $Q$. 
		The number of snapshots is $16$.
		The legend 'DPD Pilot' denotes the performance corresponding to the proposed DPD pilot design. The legend 'All-One Pilot' denotes the performance corresponding to the pilot whose elements are all one. The legend 'Upper Bound' denotes the performance upper bound of DPD scheme defined in \eqref{DPDub}.

		From Fig. \ref{DPD_Perform1}, we can observe that for different values of $N$, the proposed DPD scheme consistently achieves performance close to the theoretical upper bound, which demonstrates its effectiveness in approaching the limit of the system performance. This indicates that the DPD pilot design successfully captures the essential clutter structure and adapts the pilot to the instantaneous data payload.
		
		Furthermore, as the number of clutter components $Q$ increases, the performances of both the ``DPD Pilot'' and ``All-One Pilot'' gradually degrade. This is because a larger $Q$ introduces more intricate clutter patterns, such as the presence of richer multipath components, making it increasingly difficult for the receiver to distinguish the desired signal from clutter and noise. Consequently, the performance gap between the ``DPD Pilot'' and the ``Upper Bound'' widens with increasing $Q$, which implies that stronger clutter coupling limits the achievable SCNR despite adaptive pilot optimization.

		\subsection{DPI  Pilot Design Scheme}

		\begin{figure}[!t]
			\centering
			\includegraphics[width=3.7in]{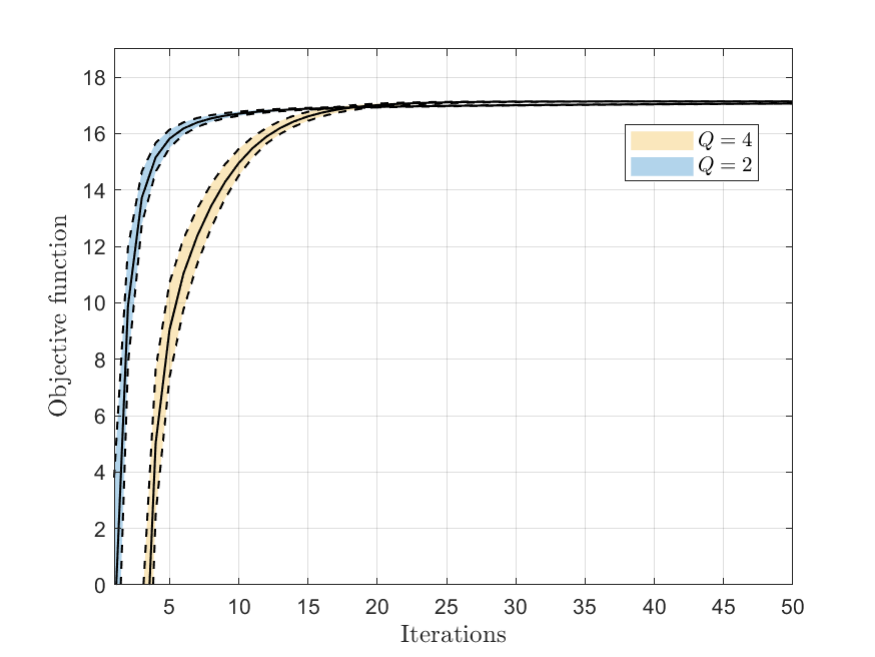}
			\caption{Convergence performance of the proposed DPI pilot design method.}
			\label{ill_DPI_Converge}
		\end{figure}
		
		\subsubsection{Convergence}
		Fig. \ref{ill_DPI_Converge} illustrates the convergence behavior of the proposed DPI pilot design algorithm. The number of snapshots is set to $N=16$. The solid line shows the average value of the objective function, while the shaded region represents its variability across realizations. As with the DPD scheme, the convergence speed is affected by the number of interferers, denoted by Q. Specifically, increasing Q makes the optimization landscape more complex and can slightly slow convergence. Nevertheless, as Fig. \ref{ill_DPD_Converge} demonstrates, the proposed method consistently converges within a few iterations for the range of Q considered, illustrating its robustness.

		\begin{figure}[!t]
			\centering
			\includegraphics[width=3.7in]{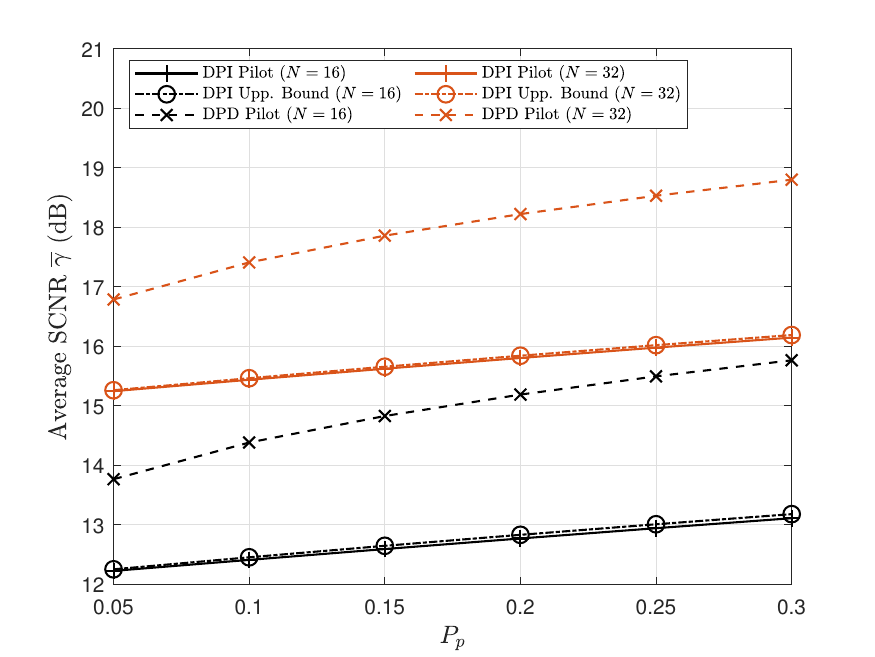}
			\caption{Performance of the proposed DPI pilot design method.}
			\label{ill_DPI_Perf}
		\end{figure}

		\subsubsection{Performance}

		Fig. \ref{ill_DPI_Perf} illustrates the variation of the average SCNR $\overline{\gamma}$ with respect to the pilot power $P_p$. The legend “DPI Pilot” and  “DPD Pilot” represent the performance achieved by the proposed DPI and DPD pilot design, respectively, while the legend ``DPI Upp. Bound'' denotes the theoretical performance limit derived in \textbf{\emph{Corollary \ref{CorUBLB}}}. It can be observed that, across different values of $P_p$, the performance of the proposed DPI pilot remains consistently close to the upper bound. This result demonstrates that the DPI pilot design effectively exploits the available pilot power to maximize the average SCNR, achieving a near-optimal performance without relying on instantaneous data payload information. Consequently, the DPI scheme provides a practical and efficient solution for pilot design with excellent robustness and power efficiency. Moreover, DPD yields better performance than DPI but requires designing a pilot for every time slot. Thus, DPI offers a practical trade‑off between complexity and performance.

		\section{Conclusion}
		
		This paper presented an AMF framework for joint sensing and communication with superimposed signals. By modeling the received waveform as a combination of deterministic pilots and random data, we derived an asymptotic expression for the average SCNR using random matrix theory, which is accurate even in moderate system dimensions. 
		Our theoretical analysis shows that, for a fixed modulation basis, PSK achieves a higher average SCNR than both QAM and Gaussian signaling. Furthermore, for any given constellation, OFDM achieves a higher average SCNR than SC and AFDM.
		Next, two pilot design schemes were proposed, i.e., the DPD and DPI methods. The DPD scheme maximizes the instantaneous SCNR for each data realization, while the DPI scheme optimizes the average SCNR to reduce computational overhead.	Simulation results verified the accuracy of the theoretical analysis and the superior sensing performance of both schemes. Future work will extend the proposed framework to multi-antenna and dynamic ISAC scenarios.

		\appendices

		\section{Proof of Proposition \ref{Prop_gammabar}}
		\label{Proof_Prop_gammabar}
		
		Note that the term $\gamma_m$ is initially defined based on an inverse of $\mathbf{R}_m$. For notational simplicity, we omit the index $m$. 
		By utilizing the Sherman Morrison formula, we have
		\begin{equation}\label{gammadef1}
			\begin{split}
				\gamma &=\frac{|\beta_0|^2 }{\sigma_n^{2} } \mathbf{x}^\HH\left(\mathbf{I} - \frac{1}{\sigma_n^2 + \mathbf{x}^\HH \mathbf{H}^\HH\mathbf{H}\mathbf{x}}\mathbf{H}\mathbf{x}\mathbf{x}^\HH \mathbf{H}^\HH\right)\mathbf{x}\\
				&=\frac{|\beta_0|^2 }{\sigma_n^{2} } \left(\mathbf{x}^\HH \mathbf{x} - \rho|\mathbf{x}^\HH\mathbf{H}\mathbf{x}|^2\right),
			\end{split}
		\end{equation}
		where 
		\begin{equation}
			\rho\triangleq\frac{1}{\sigma_n^2 + \mathbf{x}^\HH \mathbf{H}^\HH\mathbf{H}\mathbf{x}}.
		\end{equation}

		Before proceeding further, we give the following proposition:
		\begin{proposition}\label{rhoasym}
			As $N \to \infty$, we have
			\begin{equation}	
				\rho \overset{a.s.}{\to} \tilde{\rho},
			\end{equation}
			where 
			\begin{equation}
				\tilde{\rho} = \sigma_n^{-2} \left(1-  N +\sigma_n^2 \tr \mathbf{T}\right).
			\end{equation}
		\end{proposition}
		
		\textbf{\emph{Proof}}: See Appendix \ref{proofrhoasym}.  \hfill $\blacksquare$
		
		According to the Slutsky's theorem, we have
		\begin{equation}
			\begin{split}
				\gamma \overset{d}{\to} \tilde{\gamma},
			\end{split}
		\end{equation}
		where 
		\begin{equation}\label{tildegamma}
			\begin{split}
				\tilde{\gamma} =|\beta_0|^2\sigma_n^{-2}  \left(\mathbf{x}^\HH \mathbf{x} - \tilde{\rho}|\mathbf{x}^\HH\mathbf{H}\mathbf{x}|^2\right).
			\end{split}
		\end{equation}
		It implies that the random variable $\gamma$ converges in distribution to $\tilde{\gamma}$ as the dimension $N$ grows. 
		
		Thus, we have $\overline{\gamma} \triangleq \mathbb{E}(\gamma )= \mathbb{E}(\tilde{\gamma})$. The next step is to find the expectation of $\tilde{\gamma}$. To this end, we introduce the following theorem that will facilitate the derivation of the main results.
		
		\begin{proposition}\label{exp22}
			Given that $\mathbf{x}$ is a random vector with mean $\mathbb{E}(\mathbf{x})=\mathbf{x}_p$ and covariance matrix $\operatorname{Cov}(\mathbf{x})=\mathbf{I}$, we have
			\begin{subequations}
				\begin{equation}\label{exxhdef}
					\mathbb{E}\left(\mathbf{x}^\HH\mathbf{x}\right)=N+\left\Vert\mathbf{x}_p\right\Vert^2,
				\end{equation}
				\begin{equation}\label{xHx2}
					\begin{split}
						&\mathbb{E}|\mathbf{x}^\HH\mathbf{H}\mathbf{x}|^2 = \left\vert \mathbf{x}_p^\HH \mathbf{H} \mathbf{x}_p \right\vert^2 + 2\Re\left\{\left(\mathbf{x}_p^\HH \mathbf{H} \mathbf{x}_p\right) \tr \mathbf{H}^\HH\right\}\\
						& +\mathbf{x}_p^\HH \left(\mathbf{H}\mathbf{H}^\HH + \mathbf{H}^\HH\mathbf{H}\right) \mathbf{x}_p + \tr\left(\mathbf{H}\mathbf{H}^\HH\right) + \left\vert\tr\left(\mathbf{H}\right)\right\vert^2.
					\end{split}
				\end{equation}
			\end{subequations}
		\end{proposition}

		\textbf{\emph{Proof}}: See Appendix \ref{proofexp22}.  \hfill $\blacksquare$
        
By substituting \eqref{exxhdef} and \eqref{xHx2} into \eqref{tildegamma}, \eqref{overlinegamma} can be obtained.

		\section{Proof of Corollary \ref{MONO}}
		\label{proofMONO}
		
		Taking differential on $\overline{\gamma}$ with respect to $\kappa$ yields
		\begin{equation}
			\begin{split}
				\overline{\gamma}' = \frac{\partial \overline{\gamma}}{\partial \kappa} =- \frac{|\beta_0|^2}{\sigma_n^4} \tilde{\rho} \sum_{q = 1}^{Q}	\left\Vert \mathbf{b}_{n_q - n_0}\right\Vert_2^2
			\end{split}
		\end{equation}
		where
		\begin{equation}
			\begin{split}
				\tilde{\rho} = \sigma_n^{-2} \left(1-  N +\sigma_n^2 \tr \mathbf{T}\right).
			\end{split}
		\end{equation}
		From \textbf{\emph{Proposition \ref{rhoasym}}}, we can see that, in the asymptotic regime, 
		\begin{equation}
			\begin{split}
				\tilde{\rho} \overset{a.s.}{\to} \mathbb{E}(\rho) =  \mathbb{E}\left(\frac{1}{\sigma_n^2 + \mathbf{x}^\HH \mathbf{H}^\HH\mathbf{H}\mathbf{x}}\right) > 0.
			\end{split}
		\end{equation}
		Thus, we have the gradient $\overline{\gamma}'< 0$, which indicates that $\overline{\gamma}$ is monotonically decreasing with $\kappa$ in the asymptotic regime.

		\section{Proof of Proposition \ref{rhoasym}}
		\label{proofrhoasym}

		First, we have		
		\begin{equation}\nonumber
			\begin{split}
				&\rho= \sigma_n^{-2} -  \sigma_n^{-4}\mathbf{x}^\HH\mathbf{H}^\HH \left(\sigma_n^{-2}\mathbf{H}  \mathbf{x}\mathbf{x}^\HH\mathbf{H}^\HH + \mathbf{I}\right)^{-1} \mathbf{H}\mathbf{x}\\
				&= \sigma_n^{-2} \left(1-  \mathbf{x}^\HH\mathbf{H}^\HH \mathbf{\Phi} \mathbf{H}\mathbf{x}\right)\overset{(a)}{=} \sigma_n^{-2} \left[1-   \tr\left(\left(\mathbf{\Phi}^{-1} - \sigma_n^2\mathbf{I}\right) \mathbf{\Phi}\right) \right]\\
				& = \sigma_n^{-2} \left(1-  N +\sigma_n^2 \tr \mathbf{\Phi}\right),
			\end{split}
		\end{equation}
		where (a) follows from that $\mathbf{H}\mathbf{x}\mathbf{x}^\HH\mathbf{H}^\HH = \mathbf{\Phi}^{-1} - \sigma_n^2\mathbf{I}$, and
		\begin{equation}
			\begin{split}
				\mathbf{\Phi} = \left(\mathbf{H}  \mathbf{x}\mathbf{x}^\HH\mathbf{H}^\HH +\sigma_n^2 \mathbf{I}\right)^{-1}.
			\end{split}
		\end{equation}

		Since $\mathbf{\Phi}$ is a random matrix, its trace $\tr(\mathbf{\Phi})$ is a random variable, and obtaining a closed-form expression for its expectation is challenging. 
		To facilitate performance evaluation, we then derive a deterministic equivalent for $\tr( \mathbf{\Phi})$, which provides an accurate approximation in the asymptotic regime $N\to \infty$.
		
		Given the unitary invariance property of Gaussian random matrices, a Gaussian random matrix $\mathbf{G}$ is statistically equivalent to $\mathbf{G}' = \mathbf{U}' \mathbf{G} \mathbf{V}'$, for any unitary matrices $\mathbf{U}'$ and $\mathbf{V}'$. 
		Then, we have
		\begin{equation}
			\begin{split}
				\mathbf{\Phi}&=
				\left(\mathbf{U}\mathbf{\Lambda}\mathbf{V}^\HH\mathbf{x}\mathbf{x}^\HH\mathbf{V}\mathbf{\Lambda}^\HH\mathbf{U}^\HH+\sigma_n^2\mathbf{I}\right)^{-1} \\
				&=
				\mathbf{U}\left(\mathbf{\Lambda}(\mathbf{V}^\HH\mathbf{x})(\mathbf{V}^\HH\mathbf{x})^\HH\mathbf{\Lambda}^\HH+\sigma_n^2\mathbf{I}\right)^{-1}\mathbf{U}^\HH.
			\end{split}
		\end{equation}
		
		By unitary invariance, we have $\mathbf{V}^\HH \mathbf{x} \stackrel{d}{\to} \mathbf{x}$, and therefore the statistical distribution of $\mathbf{\Phi}$ coincides with that of $\mathbf{U}\mathbf{\Theta}\mathbf{U}^\HH$, where $\mathbf{\Theta} \triangleq \left(\mathbf{\Lambda}\,\mathbf{x}\mathbf{x}^\HH\mathbf{\Lambda}^\HH + \sigma_n^2 \mathbf{I}\right)^{-1}$.
		Consequently, we obtain
		\begin{equation}
			\operatorname{tr}(\mathbf{\Phi}) \overset{a.s.}{\to} \operatorname{tr}(\mathbf{\Theta}).
		\end{equation}

		Before further proceeding, we introduce the following deterministic equivalent.
		
		Consider an $N \times n$ additive matrix model $\mathbf{\Sigma}$, which is composed of a deterministic component $\mathbf{A}$ and a stochastic noise component $\mathbf{Y}$. Specifically, we define
		\begin{equation}
			\mathbf{\Sigma} = \mathbf{A} + \mathbf{Y},
		\end{equation}
		where the $(i,j)$th entry of $\mathbf{Y}$ is generated according to the model:
		\begin{equation}\label{eq:Y_model}
			[\mathbf{Y}]_{i,j} = \frac{\sigma_{i,j}}{n} [\mathbf{X}]_{i,j},  \forall i \in \{1, \dots, N\}, j \in \{1, \dots, n\}.
		\end{equation}
		In \eqref{eq:Y_model}, $\mathbf{X}$ is an $N \times n$ random matrix whose entries, $[\mathbf{X}]_{i,j}$, are i.i.d. random variables. These variables are standardized to satisfy the following statistical properties: $\mathbb{E}\left( [\mathbf{X}]_{i,j} \right) = 0$,  and  $\mathbb{E}\left( |[\mathbf{X}]_{i,j}|^2 \right) = 1.$

		Denote
		\begin{equation}
			\begin{split}
				&\mathbf{D}_j = \diag\left(|\sigma_{1,j}|^2,\cdots,|\sigma_{N,j}|^2\right),\\
				&\widetilde{\mathbf{D}}_i = \diag\left(|\sigma_{i,1}|^2,\cdots,|\sigma_{i,n}|^2\right).
			\end{split}
		\end{equation}
		Following \cite[Theorem 2.4]{hachem2007deterministic}, we can define the following fixed-point equations, i.e.,  
		\begin{equation}
			\begin{split}
				\left\{\begin{array}{l}
					\psi_i = \frac{-1}{z\left(1+(1/n)\tr\left(\widetilde{\mathbf{D}}_i  \widetilde{\mathbf{T}}\right)\right)},\; i=1,2,\cdots,N\\
					\tilde{\psi}_j = \frac{-1}{z\left( 1+(1/n)\tr\left(\mathbf{D}_j \mathbf{T}\right)\right)}, \; j=1,2,\cdots,n\\
				\end{array}
				\right.,
			\end{split}
		\end{equation}
		where
		\begin{equation}
			\begin{split}
				&\mathbf{\Psi} = \diag\left(\psi_1,\psi_2,\cdots,\psi_N\right),\\
				&\widetilde{\mathbf{\Psi}} = \diag\left(\tilde{\psi}_1,\tilde{\psi}_2,\cdots,\tilde{\psi}_n\right),\\
				&\mathbf{T} = \left(\mathbf{\Psi}^{-1} - z \mathbf{A}\widetilde{\mathbf{\Psi}} \mathbf{A}^\HH \right)^{-1},\\
				&\widetilde{\mathbf{T}} = \left(\widetilde{\mathbf{\Psi}}^{-1}-z\mathbf{A}^\HH\mathbf{\Psi}\mathbf{A}\right)^{-1}.\\
			\end{split}
		\end{equation}

		Then, we define the following resolvent matrix, i.e., 
		\begin{equation}
			\mathbf{Q} = \left(	\mathbf{\Sigma}	\mathbf{\Sigma}^\HH - z \mathbf{I}_N\right)^{-1}.
		\end{equation}
		With the above notations, the deterministic equivalent of $\mathbf{Q}$ can be formally introduced in the following theorem. 
		\begin{theorem}[{\cite[Theorem 2.5]{hachem2007deterministic}}]
			\label{TheoremRMT1}
			For given $z \in  \mathbb{C} - \mathbb{R}^{+}$, as $n \to \infty$ and $N/n \to c \in (0,\infty)$, we have	 
			\begin{equation}\label{QT}
				\frac{1}{N}\tr(\mathbf{Q}) \overset{a.s}{\to }\frac{1}{N} \tr(\mathbf{T}). 
			\end{equation}
		\end{theorem}

		By invoking $\mathbf{D} = \mathbf{\Lambda}\mathbf{\Lambda}^\HH$, $\tilde{d}_i = |\lambda_i|^2 $, $\mathbf{a} = \mathbf{\Lambda}\mathbf{x}_p$ into \textbf{\emph{Theorem \ref{TheoremRMT1}}}, we have 
		\begin{equation}\label{ThetaTasym}
			\begin{split}
				\frac{1}{N} \tr (\mathbf{\Theta}) \overset{a.s}{\to} \frac{1}{N} \tr (\mathbf{T}),
			\end{split}
		\end{equation}
		where $\mathbf{T}$ can be obtained by solving the following fixed-point equations:
		\begin{equation}\label{Tfp_eq2}
			\begin{split}
				\left\{\begin{array}{l}
					\psi_i = \frac{1}{\sigma_n^2(1+|\lambda_i|^2 \tilde{t})} ,\; i=1,2,\cdots,N,\\
					\tilde{\psi} = \frac{1}{\sigma_n^2\left( 1+\tr(\mathbf{\Lambda}\mathbf{\Lambda}^\HH \mathbf{T})\right)},\\
				\end{array}
				\right.
			\end{split}
		\end{equation}
		with
		\begin{equation}
			\begin{split}
				&\mathbf{\Psi} = \diag\left(\psi_1,\psi_2,\cdots,\psi_N\right),\\
				&\mathbf{T} = \left(\mathbf{\Psi}^{-1} + \sigma_n^2 \tilde{\psi} \mathbf{\Lambda}\mathbf{x}_p\mathbf{x}_p^\HH\mathbf{\Lambda}^\HH\right)^{-1},\\
				&\tilde{t} = \left(\tilde{\psi}^{-1} + \sigma_n^2\mathbf{x}_p^\HH\mathbf{\Lambda}^\HH \mathbf{\Psi}\mathbf{\Lambda}\mathbf{x}_p\right)^{-1}.\\
			\end{split}
		\end{equation}
		The fixed-point equations in \eqref{Tfp_eq2} are equivalent to those in \eqref{Tfp_eq}.

		\begin{remark}
			Though \textbf{\emph{Theorem \ref{TheoremRMT1}}} is rigorously derived in the asymptotic regime where $N, n \to \infty$, the resulting deterministic equivalent remains remarkably accurate for finite systems in practice \cite{xie2023sensing}. 
			As will be demonstrated numerically in Sec. V, this approximation in \eqref{ThetaTasym} is accurate in the case of moderately large $N$.
		\end{remark}
		
		\section{Proof of Proposition \ref{exp22}}
		\label{proofexp22}
		At the beginning, we give the following lemmas:
		\begin{lemma}\label{LemmaExp}
			Let $\mathbf{s} \in \mathbb{C}^{N \times 1}$ be a zero-mean complex random vector with covariance matrix $\operatorname{Cov}(\mathbf{s}) = \mathbf{\Sigma}$. Consider any deterministic positive definite matrix $\mathbf{A}$, a deterministic matrix $\mathbf{B}$, and a deterministic vector $\mathbf{a} \in \mathbb{C}^{N \times 1}$. Then, the following expectations hold:
			\begin{equation}\label{proper1}
				\mathbb{E}\left(\mathbf{a}^\HH \mathbf{s}\right) = \mathbf{a}^\HH \mathbb{E}(\mathbf{s}) = 0,
			\end{equation}
			\begin{equation}\label{proper2}
				\mathbb{E}\left(\mathbf{s}^\HH \mathbf{B} \mathbf{s}\right) = \tr\left(\mathbf{B} \mathbf{\Sigma}\right),
			\end{equation}
			\begin{equation}\label{proper3}
				\mathbb{E}\left(\mathbf{s}^\HH \mathbf{A} \mathbf{s} \cdot \mathbf{a}^\HH \mathbf{s}\right)
				= \mathbf{a}^\HH \mathbb{E}\left(\mathbf{s} \mathbf{s}^\HH \mathbf{A} \mathbf{s}\right) = 0.
			\end{equation}
		\end{lemma}
		
		\textbf{\emph{Proof}}: 1) Eq. \eqref{proper1}: 
		Using the linearity of expectation:
		\begin{equation}
			\mathbb{E}\left(\mathbf{a}^\HH \mathbf{s}\right) = \mathbf{a}^\HH \mathbb{E}(\mathbf{s}) = 0.
		\end{equation}
		
		2) Eq. \eqref{proper2}:
		Since $\mathbf{s}^\HH \mathbf{A} \mathbf{s}$ is a scalar, it is equal to its own trace. Using the cyclic property of the trace operator:
		\begin{equation}
			\begin{split}
				\mathbb{E}\left(\mathbf{s}^\HH \mathbf{A} \mathbf{s}\right) = \tr\left(\mathbf{A} \mathbb{E}(\mathbf{s} \mathbf{s}^\HH)\right) = \tr(\mathbf{A} \mathbf{\Sigma}).
			\end{split}
		\end{equation}
		
		3) Eq. \eqref{proper3}:
		The expression \eqref{proper3} involves third-order moments of the elements of $\mathbf{s}$. For a circularly-symmetric distribution, all odd-order moments are zero. Therefore, the expectation is zero. \hfill $\blacksquare$

		\begin{lemma}[{\cite[Proposition 1]{liu2025cp}}]\label{SLexp}
			Let $\mathbf{s}$ be a random vector which follows \textbf{\emph{Assumption \ref{Asmp1}}}. 
			For all $k\neq 0$, we have
			\begin{equation}\label{Ezbz4}
				\begin{split}
					\mathbb{E}\left( \left\vert \mathbf{s}^\HH\mathbf{U}^\HH\mathbf{J}_{k}\mathbf{U}\mathbf{s} \right\vert^2 \right) = N + (\kappa - 2)\left\Vert \mathbf{b}_k\right\Vert_2^2.
				\end{split}
			\end{equation}
		\end{lemma}

		Next, we proceed to the proof of \textbf{\emph{Proposition \ref{exp22}}}. 
		By taking expectations and using linearity of expectation and the cyclic property of the trace, we have
		\begin{equation}\label{exxH}
			\begin{split}
				\mathbb{E}(\mathbf{x}^\HH \mathbf{x})&=\tr\left(\mathbb{E}[\mathbf{x}\mathbf{x}^\HH]\right)
				=\tr\left(\operatorname{Cov}(\mathbf{x})+\mathbf{x}_p\mathbf{x}_p^\HH\right)
			\end{split}
		\end{equation}
		By utilizing \eqref{proper2}, we have
		\begin{equation}\label{Convx}
			\tr(\operatorname{Cov}(\mathbf{x})) = \tr(\mathbf{I}) = N.
		\end{equation}
		By substituting \eqref{Convx} into \eqref{exxH}, \eqref{exxhdef} can be obtained.

		Next, we observe that
		\begin{equation}\label{xhx}
			\begin{split}
				&\left\vert\mathbf{x}^\HH\mathbf{H}\mathbf{x}\right\vert^2 \\
				&= 
				\left\vert\left(\mathbf{x}_p + \mathbf{U}\mathbf{s}_d\right)^\HH\mathbf{H}\left(\mathbf{x}_p + \mathbf{U}\mathbf{s}_d\right)\right\vert^2  = \left\vert
				a + b + c + d \right\vert^2\\
				& = \underbrace{aa^*}_{\text{\ding{172}}} + \underbrace{bb^*}_{\text{\ding{173}}} + \underbrace{cc^*}_{\text{\ding{174}}} + \underbrace{dd^*}_{\text{\ding{175}}} + \underbrace{2\Re(ba^*)}_{\text{\ding{176}}} + \underbrace{2\Re(ca^*)}_{\text{\ding{177}}} \\
				&\quad + \underbrace{2\Re(da^*)}_{\text{\ding{178}}} + \underbrace{2\Re(cb^*)}_{\text{\ding{179}}} + \underbrace{2\Re(db^*)}_{\text{\ding{180}}} + \underbrace{2\Re(dc^*)}_{\text{\ding{181}}},
			\end{split}
		\end{equation}
		where
		\begin{equation}
			\begin{split}
				&a = \mathbf{x}_p^\HH \mathbf{H} \mathbf{x}_p,\; b= \mathbf{x}_p^\HH \mathbf{H} \mathbf{U}\mathbf{s}_d,\\
				&c = \mathbf{s}_d^\HH \mathbf{U}^\HH \mathbf{H} \mathbf{x}_p,\; d = \mathbf{s}_d^\HH \mathbf{U}^\HH \mathbf{H} \mathbf{U} \mathbf{s}_d.
			\end{split}
		\end{equation}

		Since the term \ding{172} is deterministic, we give the expectation for the terms \ding{173}-\ding{181} as follows:

		\subsubsection{Terms \ding{177} and \ding{178}}
		By utilizing the expectation in \eqref{proper1}, we have
		\begin{equation}
			\begin{split}
				ba^* &= \mathbb{E}\left( \mathbf{x}_p^\HH \mathbf{H}^\HH \mathbf{x}_p  \mathbf{x}_p^\HH \mathbf{H} \mathbf{U}\mathbf{s}_d \right)  = 0,
			\end{split}
		\end{equation}
		\begin{equation}
			\begin{split}
				c a^* &= \mathbb{E}\left( \mathbf{x}_p^\HH \mathbf{H}^\HH \mathbf{x}_p \mathbf{s}_d^\HH \mathbf{U}^\HH \mathbf{H} \mathbf{x}_p \right)  = 0.
			\end{split}
		\end{equation}

		\subsubsection{Terms \ding{173}, \ding{174}, and \ding{178}}
		By utilizing the expectation in \eqref{proper2}, we have
		\begin{equation}\nonumber
			\begin{split}
				bb^* &= \mathbb{E}\left(\mathbf{s}_d^\HH \mathbf{U}^\HH \mathbf{H}^\HH \mathbf{x}_p\mathbf{x}_p^\HH \mathbf{H} \mathbf{U}\mathbf{s}_d \right) = \tr(\mathbf{U}^\HH \mathbf{H}^\HH \mathbf{x}_p\mathbf{x}_p^\HH \mathbf{H} \mathbf{U})\\
				& = \mathbf{x}_p^\HH \mathbf{H}\mathbf{U} \mathbf{U}^\HH\mathbf{H}^\HH \mathbf{x}_p,
			\end{split}
		\end{equation}
		\begin{equation}\nonumber
			\begin{split}
				cc^* &= \mathbb{E}\left(\mathbf{s}_d^\HH \mathbf{U}^\HH \mathbf{H} \mathbf{x}_p\mathbf{x}_p^\HH \mathbf{H}^\HH \mathbf{U}\mathbf{s}_d \right) = \tr(\mathbf{U}^\HH \mathbf{H}  \mathbf{x}_p\mathbf{x}_p^\HH \mathbf{H}^\HH \mathbf{U})\\
				& = \mathbf{x}_p^\HH \mathbf{H}^\HH\mathbf{U} \mathbf{U}^\HH\mathbf{H} \mathbf{x}_p,
			\end{split}
		\end{equation}
		\begin{equation}\nonumber
			\begin{split}
				d a^* &= \mathbb{E}\left( \mathbf{x}_p^\HH \mathbf{H}^\HH \mathbf{x}_p \mathbf{s}_d^\HH \mathbf{U}^\HH \mathbf{H} \mathbf{U} \mathbf{s}_d \right)  = \mathbf{x}_p^\HH \mathbf{H}^\HH \mathbf{x}_p \tr \mathbf{H} = 0.
			\end{split}
		\end{equation}

		\subsubsection{Term \ding{180} and \ding{181}} 
		By utilizing the expectation in \eqref{proper3}, we have
		\begin{equation}
			\begin{split}
				d b^* &= \mathbb{E}\left( 
				(\mathbf{s}_d^\HH \mathbf{U}^\HH \mathbf{H} \mathbf{U} \mathbf{s}_d )(\mathbf{x}_p^\HH \mathbf{H} \mathbf{U}\mathbf{s}_d)^*
				\right)  = 0,
			\end{split}
		\end{equation}
		\begin{equation}
			\begin{split}
				d c^* &= \mathbb{E}\left( 
				(\mathbf{s}_d^\HH \mathbf{U}^\HH \mathbf{H} \mathbf{U} \mathbf{s}_d )(\mathbf{s}_d^\HH \mathbf{U}^\HH \mathbf{H} \mathbf{x}_p)^*
				\right)  = 0.
			\end{split}
		\end{equation}

		\subsubsection{Term \ding{175}} 
		Let $\mathbf{z}$ be a random vector which follows \textbf{\emph{Assumption \ref{Asmp1}}}.	From the definition of $\mathbf{H}$, we have
		\begin{equation}\label{Ezbz}
			\begin{split}
				&\mathbb{E}\left(|\mathbf{z}^\HH \mathbf{H} \mathbf{z}|^2\right) = \mathbb{E}\left(\left\vert\mathbf{z}^\HH \left(\sum_{q=1}^Q \beta_q \mathbf{J}_{n_q - n_0} \right) \mathbf{z}\right\vert^2\right) \\
				& = \mathbb{E}\left[\left(\sum_{q=1}^Q \beta_q \mathbf{z}^\HH\mathbf{J}_{n_q - n_0}\mathbf{z} \right)\left(\sum_{p=1}^Q \beta_p \mathbf{z}^\HH\mathbf{J}_{n_p - n_0}\mathbf{z} \right)^\HH\right]\\
				& = \underbrace{\mathbb{E}\left(\sum_{q=1}^Q |\beta_q|^2 \left\vert \mathbf{z}^\HH\mathbf{J}_{n_q - n_0}\mathbf{z} \right\vert^2 \right)}_{\mathrm{Auto \;Term}} \\
				&\quad + 
				\underbrace{\mathbb{E}\left( \sum_{p\neq q}\beta_q\beta_p^* \mathbf{z}^\HH\mathbf{J}_{n_q - n_0}\mathbf{z}\left(\mathbf{z}^\HH\mathbf{J}_{n_p - n_0}\mathbf{z}\right)^*\right)}_{\mathrm{Cross \;Term}}.
			\end{split}
		\end{equation}
		
		The two components of the summation in \eqref{Ezbz} will now be analyzed separately:
		
		1) Cross Term:
		For all $q \neq p$, we have
		\begin{equation}\label{Ezbz2_eng}
			\begin{split}
				&\mathbb{E}\left[\left(\mathbf{z}^\HH\mathbf{J}_{q}\mathbf{z}\right)\left(\mathbf{z}^\HH\mathbf{J}_{p}\mathbf{z}\right)^* \right] \\
				&= \mathbb{E}\left[\left(\sum_{i = 1}^{N} z_i^* z_{i+q} \right) \left(\sum_{j = 1}^{N} z_j^* z_{j+p} \right)^* \right] \\
				&= \mathbb{E}\left[\left(\sum_{i = 1}^{N} z_i^* z_{i+q} \right) \left(\sum_{j = 1}^{N} z_j z_{j+p}^* \right) \right] \\
				&= \sum_{i = 1}^{N}\sum_{j = 1}^{N} \mathbb{E}\left[z_i^* z_{i+q} z_j z_{j+p}^*\right] = 0.
			\end{split}
		\end{equation}
		We now elaborate on the reasoning for the final equality.
		
		Note that the components $z_k, k=1, \dots, N$ of the random vector $\mathbf{z}$ are i.i.d. complex random variables with zero mean and circular symmetry. Let us now examine the expectation term $\mathbb{E}\left[z_i^* z_{i+q} z_j z_{j+p}^*\right]$. For this expectation to be non-zero, the four variables must form two conjugate pairs. This requires the indices of the variables to match accordingly. Specifically, the set of indices of the non-conjugated variables, $\{i+q, j\}$, must be a permutation of the set of indices of the conjugated variables, $\{i, j+p\}$. This leads to two possible scenarios:
		\begin{itemize}
			\item \emph{Case 1:} $i+q = i$ and $j = j+p$. \\
			This would require $q=0$ and $p=0$. This contradicts the context of using shift matrices (where $p, q$ represent non-zero delays) and also violates the premise that $q \neq p$.
			
			\item \emph{Case 2:} $i+q = j+p$ and $j = i$. \\
			Substituting $j=i$ into the first equation gives $i+q = i+p$, which directly implies $q=p$. However, this contradicts our fundamental assumption that $q \neq p$.
		\end{itemize}
		
		Since neither of the two scenarios that could yield a non-zero expectation can be satisfied under the condition $q \neq p$, the expectation term $\mathbb{E}\left[z_i^* z_{i+q} z_j z_{j+p}^*\right]$ must be zero for all combinations of $i$ and $j$.
		
		Consequently, as every term in the double summation is zero, the entire sum is zero. This proves that:
		\begin{equation}
			\mathbb{E}\left[\left(\mathbf{z}^\HH\mathbf{J}_{q}\mathbf{z}\right)\left(\mathbf{z}^\HH\mathbf{J}_{p}\mathbf{z}\right)^* \right] = 0, \quad \forall q \neq p.
		\end{equation}

		2) Auto Term: Therefore, from \eqref{Ezbz}, we have
		\begin{equation}\label{Ezbz3}
			\begin{split}
				&\mathbb{E}\left(|\mathbf{s}_d^\HH \mathbf{U}^\HH\mathbf{H} \mathbf{U}\mathbf{s}_d|^2\right) = \sum_{q=1}^Q |\beta_q|^2\mathbb{E}\left( \left\vert \mathbf{s}_d^\HH \mathbf{U}^\HH\mathbf{J}_{n_q - n_0}\mathbf{U}\mathbf{s}_d\right\vert^2 \right) .
			\end{split}
		\end{equation}
		By utilizing \textbf{\emph{Lemma \ref{SLexp}}}, we have
		\begin{equation}
			\begin{split}
				d d^* &= 	\sum_{q=1}^Q |\beta_q|^2\left(N + (\kappa - 2)\left\Vert \mathbf{b}_{n_q-n_0}\right\Vert_2^2\right).
			\end{split}
		\end{equation}

		\subsubsection{Term \ding{179}} 
		By the property of circular symmetry, we have
		\begin{equation}
			\begin{split}
				c b^* &= \mathbb{E}\left( 
				(\mathbf{s}_d^\HH \mathbf{U}^\HH \mathbf{H} \mathbf{x}_p )(\mathbf{x}_p^\HH \mathbf{H} \mathbf{U}\mathbf{s}_d)^*
				\right)  = 0.
			\end{split}
		\end{equation}
		
		By substituting \ding{173}-\ding{181} into \eqref{xhx}, \eqref{xHx2} can be obtained.


	\end{document}